\documentclass[preprint,showpacs,preprintnumbers,amsmath,amssymb]{revtex4}

\usepackage{graphicx}% Include figure files
\usepackage{dcolumn}% Align table columns on decimal point
\usepackage{bm}% bold math
\usepackage{amssymb}
\usepackage{amsmath}
\usepackage{times}
\usepackage{epsfig}
\usepackage[english]{babel}

\begin{document}

\title{2D pattern evolution constrained by complex network dynamics}

\author{L. E. C. da Rocha and L. da F. Costa}
\email{luciano@if.sc.usp.br}

\affiliation{Grupo de Pesquisa em Vis\~ao Cibern\'etica, 
Instituto de F\'isica de S\~ao Carlos,
Universidade de S\~ao Paulo, Av. Trabalhador S\~ao Carlense 400,
Caixa Postal 369, 13560-970, S\~ao Carlos, SP, Brazil}

\date{29th September 2006}

\begin{abstract}

Complex networks have established themselves along the last years as
being particularly suitable and flexible for representing and
modeling several complex natural and human-made systems.  At the
same time in which the structural intricacies of such networks are
being revealed and understood, efforts have also been directed at
investigating how such connectivity properties define and constrain
the dynamics of systems unfolding on such structures. However,
lesser attention has been focused on hybrid systems, \textit{i.e.}
involving more than one type of network and/or dynamics.  Because
several real systems present such an organization (\textit{e.g.} the
dynamics of a disease coexisting with the dynamics of the immune
system), it becomes important to address such hybrid systems.  The
current paper investigates a specific system involving a diffusive
(linear and non-linear) dynamics taking place in a regular network while interacting with
a complex network of defensive agents following Erd\"os-R\'enyi and Barab\'asi-Albert
graph models, whose nodes can be displaced spatially.  More
specifically, the complex network is expected to control, and if
possible to extinguish, the diffusion of some given unwanted process
(\textit{e.g.} fire, oil spilling, pest dissemination, and virus or
bacteria reproduction during an infection). Two types of pattern
evolution are considered: Fick and Gray-Scott. The nodes of the defensive 
network then interact with the diffusing patterns and communicate between 
themselves in order to control the spreading. The main
findings include the identification of higher efficiency 
for the Barab\'asi-Albert control networks.
\end{abstract}


%\pacs{89.75.Hc}{Networks and Genealogical Trees}
%\pacs{89.75.Kd}{Patterns (complex systems)}
%\pacs{05.45.-a}{Nonlinear dynamics and chaos}
%\pacs{02.70.Rr}{General Statistical Methods}

\maketitle

\section{Introduction}

Complex systems have always motivated intense scientific research. In
the last decades, much attention has been focused on systems involving
strongly interacting agents. More recently, tools provided by the
theory of complex networks have been successfully applied in order to
characterize the structure of many of such
systems~\cite{Newman_review, Dorog_review, BaraAlbert_review}. Once
the system of interest is properly translated into a network, its
structural properties~\cite{Dorog_book,Costa_survey, Newman_review, BaraAlbert_review} can be calculated
and used to characterize and analyze the system as well as dynamical
processes being underlined by the network~\cite{Motter_cascades,
Madar_epidemic, GallosArgyrakis_readif, NohRieger_randwalk, Newman_review}. However,
many dynamics are often related to processes taking place outside the
network, possibly also over some network (the same or different).
Such systems have received scant attention from the complex network
community.

The current paper investigates the evolution of dynamical
systems underlined by two distinct (but coexisting) networks, which
are henceforth called \textit{disease} and \textit{antidote}.  Note that this
specific terminology is adopted here only for the sake of
simplicity; the proposed model and dynamics are valid for 
many situations (\textit{e.g.} fire spread, oil
spilling, pest control, etc.) other than diseases
and inflammatory processes. The first system, involving a complex
network of the Erd\"os-R\'enyi -- ER~\cite{ErdoesRenyi_model} or
Barab\'asi-Albert -- BA~\cite{BaraAlbert_model} type, senses and
interact with the other system, here represented by a \textit{regular
network} over which linear (Fick~\cite{Crank_book}) and non-linear
(Gray-Scott~\cite{GrayScott_model}) pattern formation is allowed to
evolve. The Fick diffusion model provides a linear,
homogeneous and isotropic flux of mass from a fixed and infinite
source. The Gray-Scott reaction-diffusion dynamics produces
non-static, growing patterns without well-defined sources. Examples
of such situations include forest fires, where the nodes of the
complex networks represent firemen, organized into communicating
groups, trying to stop the spreading of the fire, represented by a
diffusive process in the regular network. Other similar situations
include oil spilling (oil diffusing along the regular network, while
a complex network of cleaners try to control the process) and the
evolution of a disease along a healthy tissue, with the nodes
representing the defensive cells trying to self-organize in order to
control and stop the disease. Observe that the connections estabilished by the agents of the system are not necessarily physical. In fact, these connections may correspond to wireless communication, bio-chemical signaling or even intermediate agents (as modeled in \textit{bi-partite graphs}), \textit{e.g.} enzymes in biological networks.

The article starts by presenting the pattern formation models (Fick
and Gray-Scott) and proceeds by describing the interaction between the
two involved networks (\textit{i.e.} \textit{regular} and \textit{complex}).  
The results and discussion follow, and the article is conclude by
emphasizing the main contributions and perspectives for future
developments.

\section{Diffusion models}

Over the twentieth century, a number of natural phenomena have been
modeled by diffusion and pattern formation processes. The former
type of dynamics includes the established topics of atoms and
molecules diffusion~\cite{Carslaw_book} as well as heat diffusion
through different materials ~\cite{Kadanoff_book}. In addition,
econometricians have developed diffusion models to forecast the
acceptance of new products and the understanding of their
life-cycle~\cite{Wilmott_book}. Migration of animals and spreading of organisms and chemical substances are
often investigated in terms of biological diffusion
models~\cite{Okubo_book}. More recently, complex biological and
chemical patterns have been reproduced by systems of equations with
diffusive and reactive terms~\cite{Grindrod_book}. These models
range from simple diffusion equations (\textit{e.g.} heat diffusion
in a rod) to more sophisticated advection-diffusion (\textit{e.g.},
chemical oceanography) and reaction-diffusion equations
(\textit{e.g.}, chemical and biological patterns). Two of such
models are considered in the present paper in order to represent a
reasonably representative range of natural and artificial phenomena:
Fick diffusion and the Gray-Scott reaction-diffusion models.

The Fick diffusion model of an entity $U$ is represented in
eq.~(\ref{eq:01}). It can be derived from the continuity equation
~\cite{Crank_book}. The concentration $u$ of $U$ evolve in time
proportionally to the difference between the average value of $u$
around a given point and the value of $u$ at that point. The
proportionality constant is given by the diffusion coefficient
$D_{u}$.

\begin{equation}
\label{eq:01}
\frac{\partial u}{\partial t} = D_{u}\nabla^{2}u
\end{equation}

The Gray-Scott model includes the following two irreversible reactions
~\cite{GrayScott_model}:
\begin{equation}
\label{eq:02}
U + 2V \rightarrow 3V
\end{equation}
\begin{displaymath}
V \rightarrow P
\end{displaymath}

where $U$ and $V$ are two reacting specimens and $P$ an inert
precipitate. Considering the concentrations of specimens $U$ and $V$,
respectively as $u$ and $v$, these reactions can be expressed by a pair of non-linear partial differential
equations~(\ref{eq:03}) with diffusive and reactive terms.

\begin{equation}
\label{eq:03}
\frac{\partial u}{\partial t} = D_{u}\nabla^{2}u - uv^2 + f(1-u)
\end{equation}
\begin{displaymath}
\frac{\partial v}{\partial t} = D_{v}\nabla^{2}v + uv^2 - (f+k)v
\end{displaymath}

where $D_{u}$ and $D_{v}$ are the diffusion coefficients. The dimensionless
feed rate of the first reaction is $f$; $k$ is the dimensionless rate
constant of the second reaction.

\section{Diffusion and defense dynamics}

Both diffusion models were evaluated on a spatial mesh
(\textit{i.e.}, a regular network) of $256$ by $256$ points with
periodic boundary conditions. The system size was $3.0$ in both
directions. Numerical integrations were carried out by the forward
Euler method of the finite-difference equations resulted from
discretization of the diffusion operator. The time step was $1$ time
unit. The diffusion coefficients were set as $D_{u}=0.00002$ (to
both diffusion models) and $D_{v}=0.00001$. A complex network was
used to represent the agents (\textit{i.e.}, nodes) susceptible to
be activated by the regular network. There were two states
associated to each node: \textit{susceptible} or \textit{activated}. All the nodes
began in the susceptible state. As soon as the disease overcame a
threshold at the node spatial position $(x,y)$, or in case the node is
requested to help its neigbors, the node was turned to the
activated state. In case a node is requested simultaneously as a consequence of high activity in the regular network and by one of its neigbors in the complex network, priority is given to the
former situation. After a while, the node returned to the susceptible state.

\begin{figure}[ht]
  \begin{center}
  \includegraphics[scale=0.5]{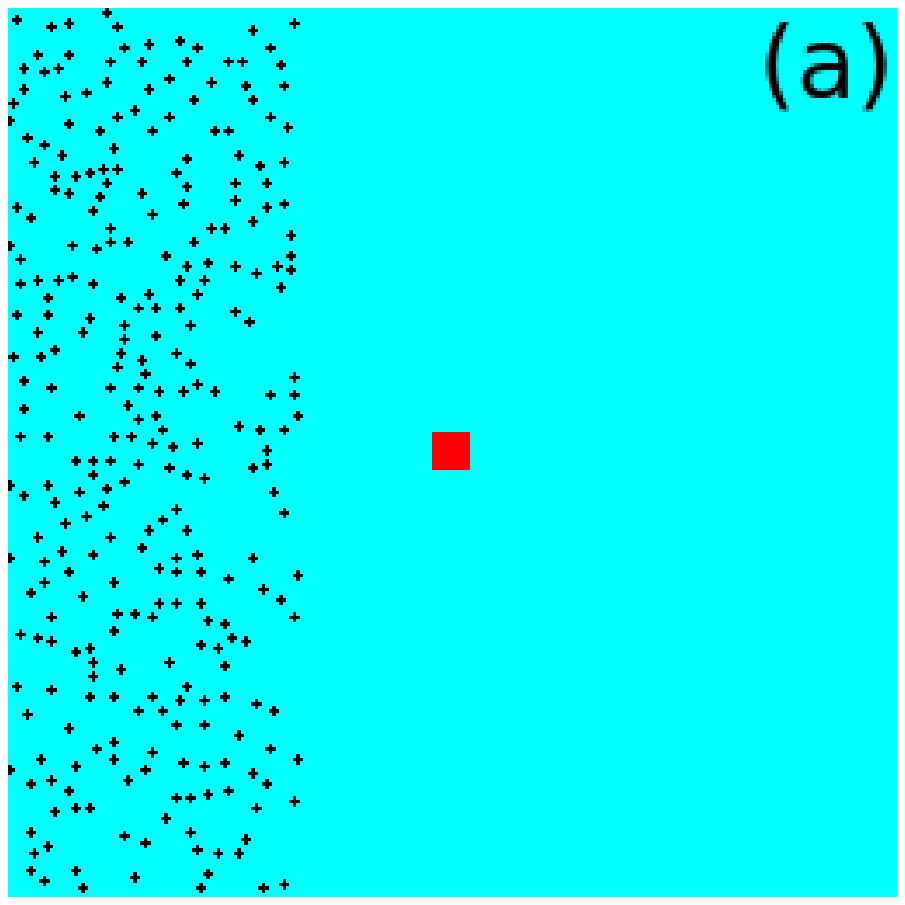}
  \includegraphics[scale=0.5]{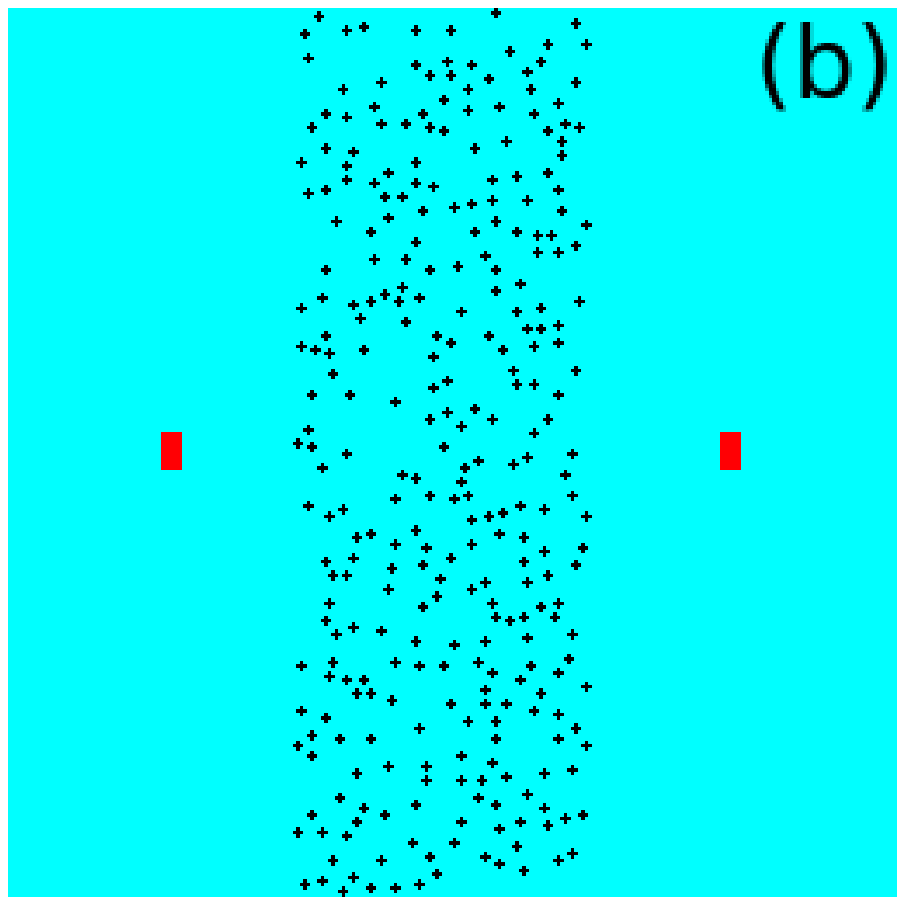}
  \caption{Two configurations of initial conditions for the Fick
  diffusion model: (a) one source and (b) two sources. Similar initial
  conditions were used for the Gray-Scott model, except for the source
  value.}
  \label{fig:01}
  \end{center}
\end{figure}

Two configurations of initial conditions were investigated. In the
first configuration (fig.~\ref{fig:01}-a), the entire system was
placed in the uninfected state: $U(x,y)=0$ (Fick model) and, $U(x,y)=1$ and
$V(x,y)=0$ (Gray-Scott model). The source of the disease, a $11$ by $11$
square mesh points, was centered in the middle of the board and set
as $U(x,y)=1$ (Fick model), and $U(x,y)=0.5$ and $V(x,y)=0.25$ (Gray-Scott model). In the latter model, the source was perturbed by adding random values 
of $\pm 0.01$, in order to break the square symmetry. The node were randomly distributed inside a rectangular area (one third
of the board area, with $256$ by $85$ points) on the left side of the mesh, at $38$ mesh points away from the disease source.
Initially, all the node were set to the susceptible state. This simple arrangement was chosen to create a \textit{"wall"} of node and contributed to the vertical symmetry of the configuration, reducing the
number of parameters to be considered during simulation.

In the second configuration (fig.~\ref{fig:01}-b), the node were
distributed inside the same rectangular region as before, but the area
was centered in the middle of the mesh. The source was broken in two
($11$ by $6$ rectangular mesh points each piece), to correspond to about the same
amount of initial disease. Both sources were symmetrically placed at
same distance (\textit{i.e.}, $38$ mesh points) and opposite sides
from the nodes \textit{"wall"}. This assembly induced a competition for
neigbors of activated node.

\begin{figure}[ht]
  \begin{center}
  \includegraphics[scale=0.3]{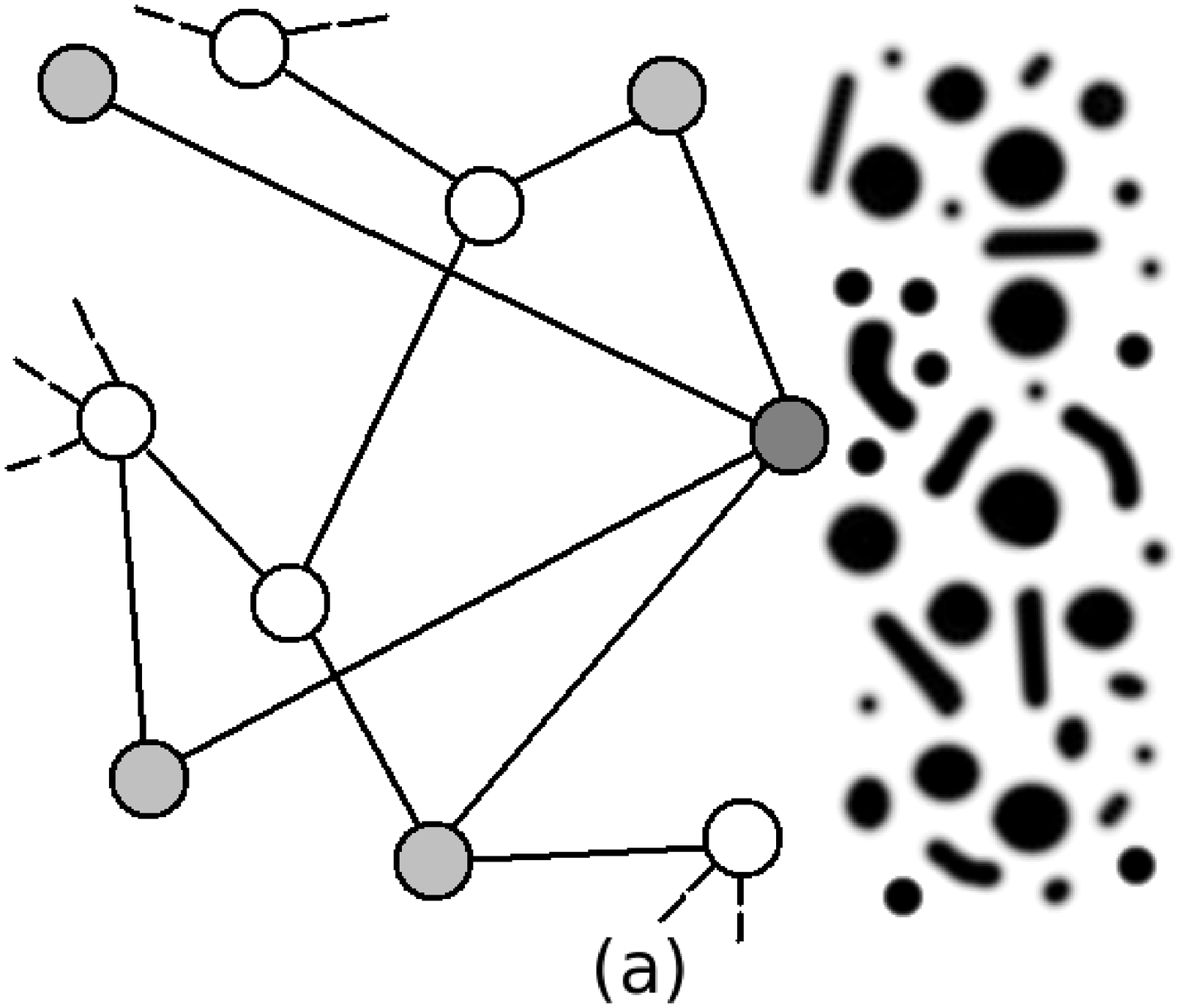}
  \includegraphics[scale=0.3]{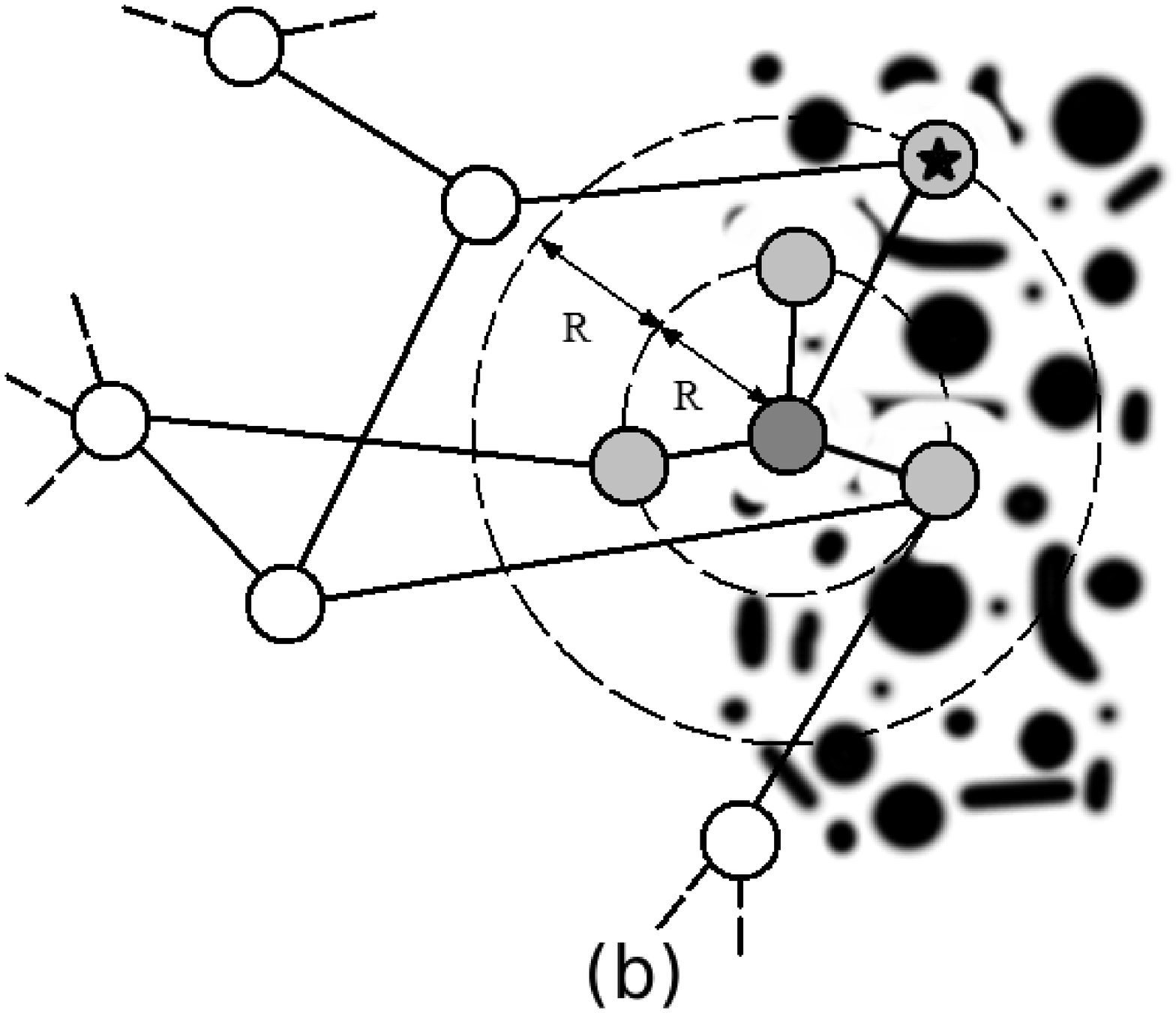}
  \caption{Diagram illustrating a \textit{sub-network} (a) before and (b)
  after the node activation. The dark-gray node was spatially activated
  by the disease and the four light-gray node were the topologically
  activated helpers.}
  \label{fig:02}
  \end{center}
\end{figure}

In the Fick model, a node became activated when the disease overcame a threshold $T_{U}(x,y)=0.4$ at the respective node position, \textit{i.e.}, \textit{x} and \textit{y}. In the
Gray-Scott model, the disease must fell below a threshold $T_{U}(x,y)=0.6$
in order to activate the node. Remember that absence of disease was
represented by $U(x,y)=0$ in the Fick model and by $U(x,y)=1$ in the Gray-Scott
model. As soon as a node had been activated, all its topological
neigbors were requested to help (see fig.~\ref{fig:02}). The engaged
neigbors were randomly distributed at distance $R=5$ fromoi the activated
node. In order to avoid overlapping in the liberation of antidote, a circular
area of influence (with radius $R_{i}=5$) was defined around every
node, so that no other activated node was included within this area. In
fact, we guaranteed a minimum distance ($R=R_{i}=5$) between any two
activated node, ensuring a compact distribution of the node. Once
this circle was filled, the remaining node were assembled at double the
initial radius, and so on (see, for example, the node with a \textit{star} in
fig.~\ref{fig:02}). The antidote liberation consisted in keeping for
$50$ time units an opposite Fick diffusion from all activated node
with $D_{a}=0.00003$, and intensity $I_{a}(x,y)=1$ (Fick model) and
$I_{a}(x,y)=10$ (Gray-Scott model). The higher intensity is necessary in
the latter model because of the fast moving characteristic of this
reaction-diffusion. Observe that the activated time is calculated so as to liberate enough antidote within the circular area of influence of the node, reducing the overlap between different node. Afterwards,
the node ceased its activity and returned to the susceptible state. If two node requested help from the same neighbor, the latter chose one of them with equal probability.

\section{Results}

The suggested dynamics involving the interaction between two networks
always resulted in competition between the disease and the antidote,
where the winner was ultimately a consequence of the values chosen for
the diffusion and defense parameters. Some parameter configurations
have been observed to lead to a situation where great part of the
effort to control the disease was wasted. On the other hand, it was possible to find parameter configurations where the defense always succeed, \textit{i.e}, the disease vanished. Once such
parameters were identified and adopted, we compared the role of the
network structure in the proposed defense dynamics.

\begin{figure*}
  \begin{center}
    (a) \\
    \includegraphics[scale=0.33]{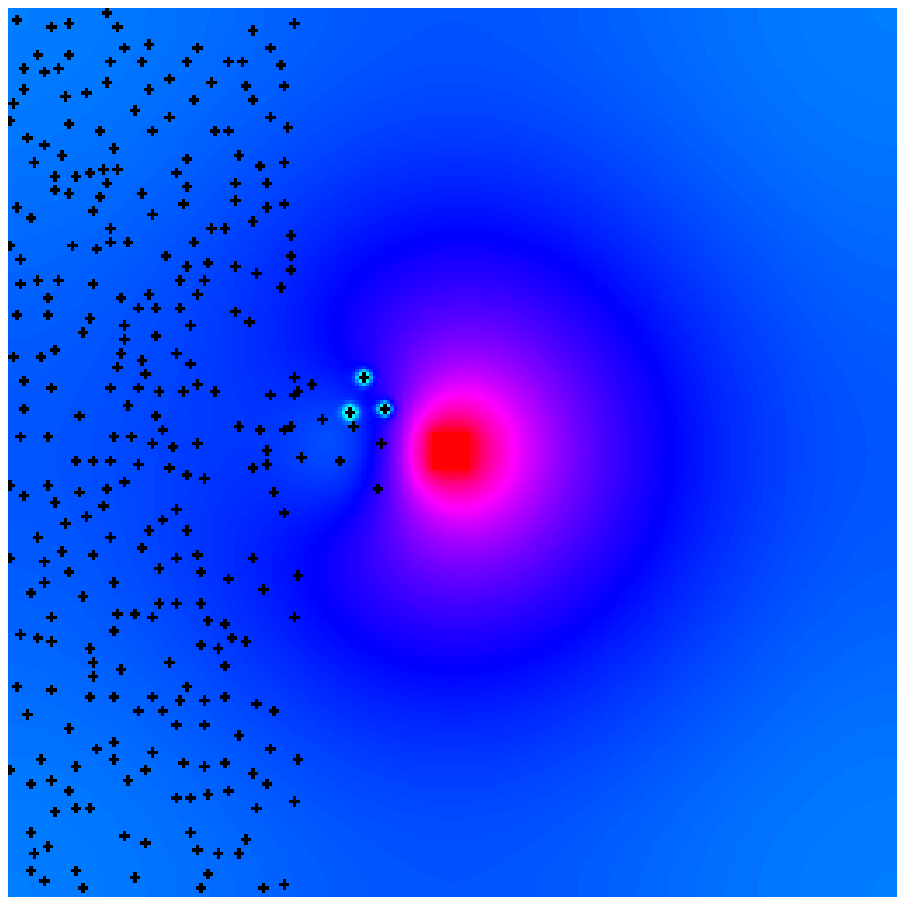}
    \includegraphics[scale=0.33]{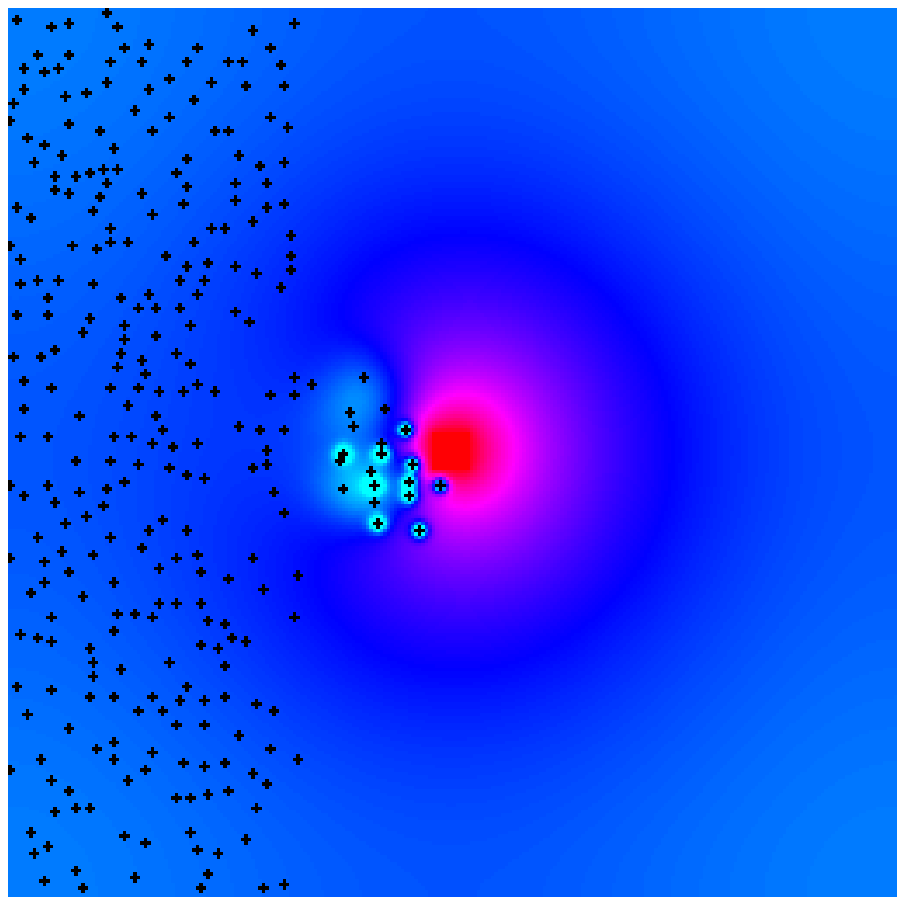}
    \includegraphics[scale=0.33]{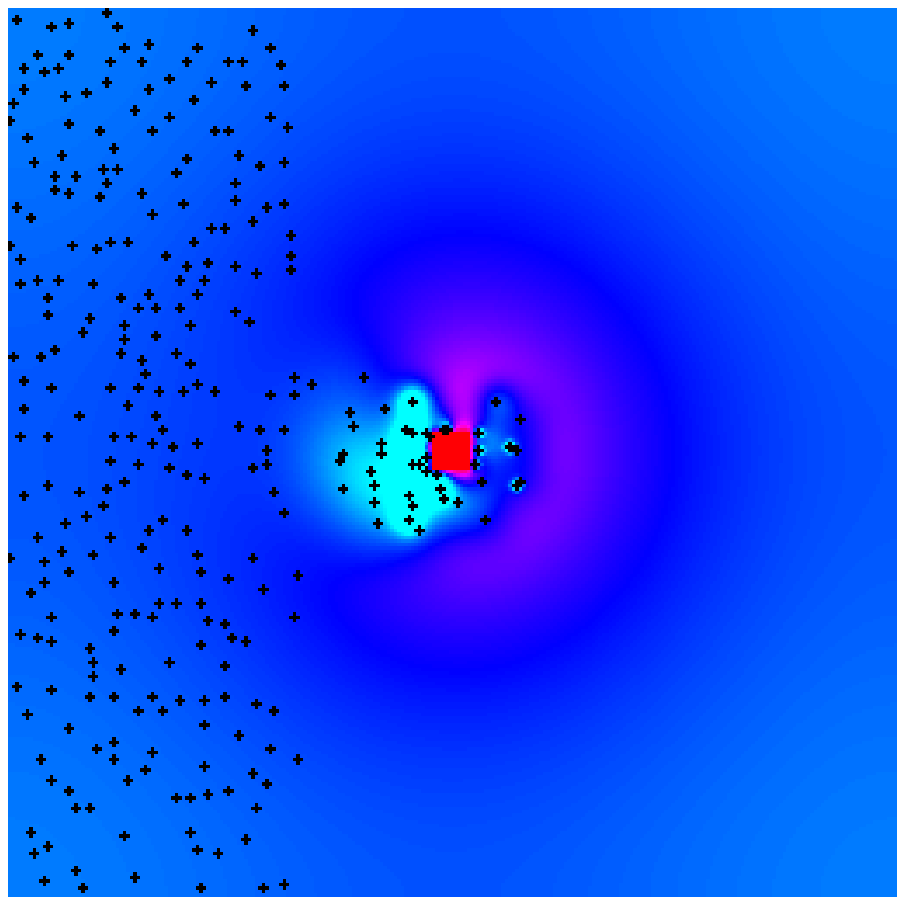}
    \includegraphics[scale=0.33]{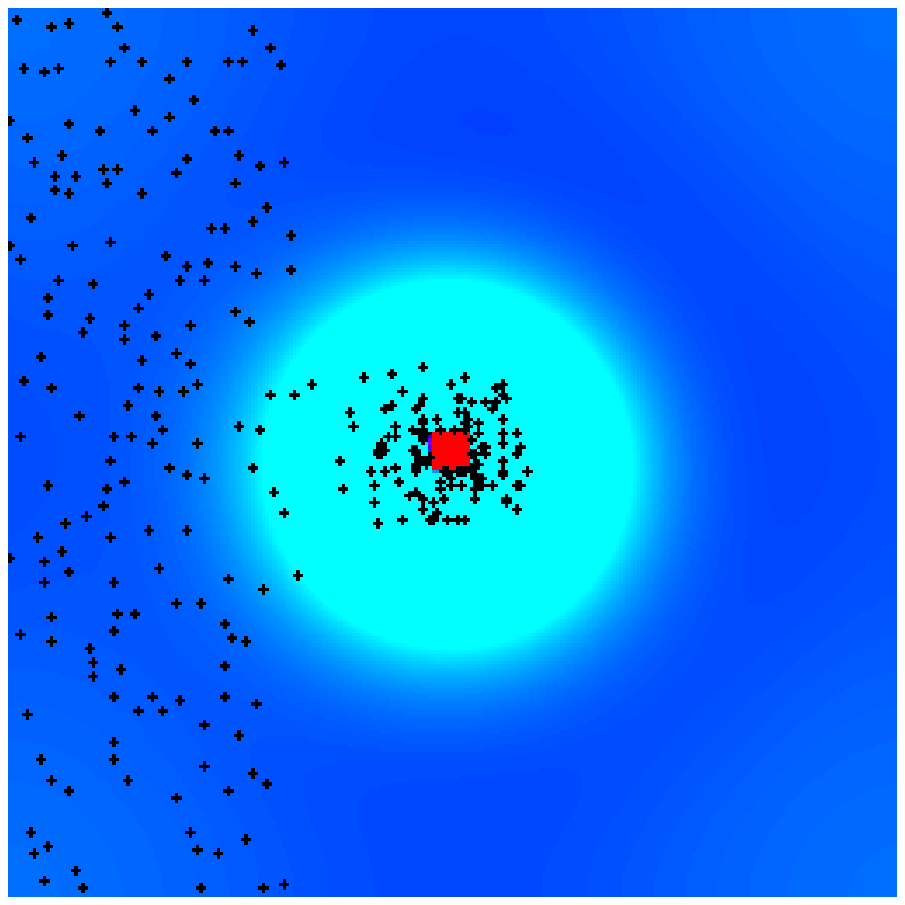}
    \includegraphics[scale=0.33]{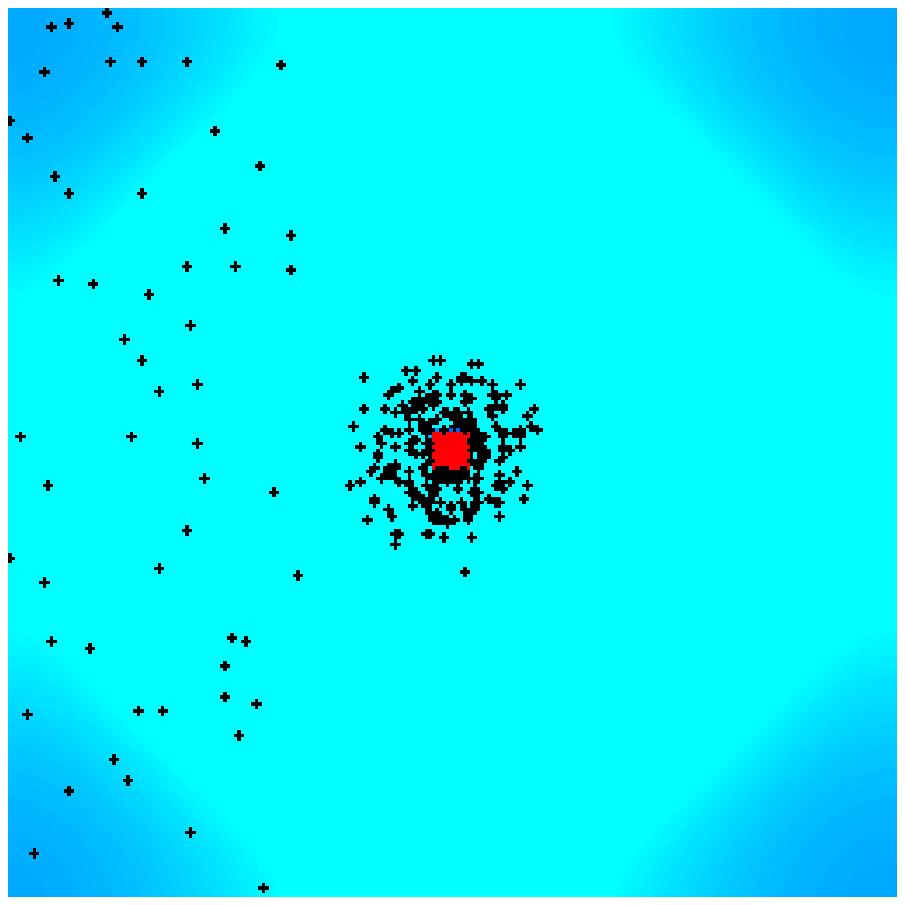}   \\
    $t=55104$ \hspace{1.3cm}   $t=55304$ \hspace{1.3cm} $t=55604$ \hspace{1.3cm} $t=60000$ \hspace{1.3cm} $t=80000$ \\
    (b) \\
    \includegraphics[scale=0.33]{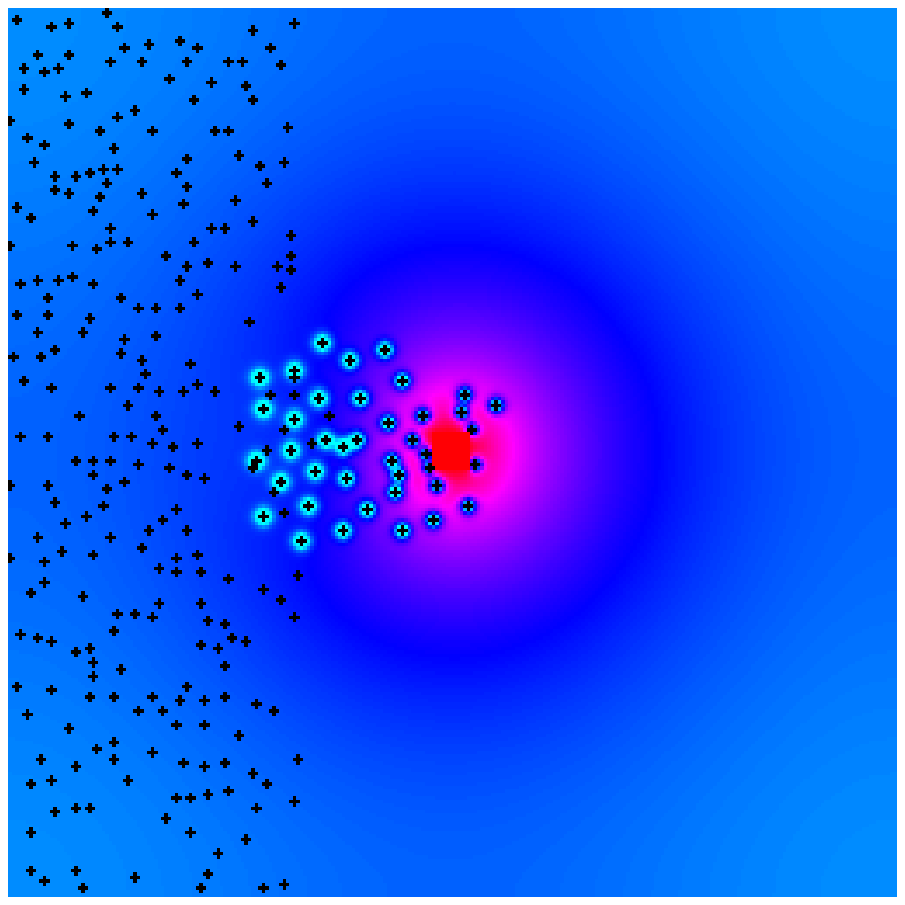}
    \includegraphics[scale=0.33]{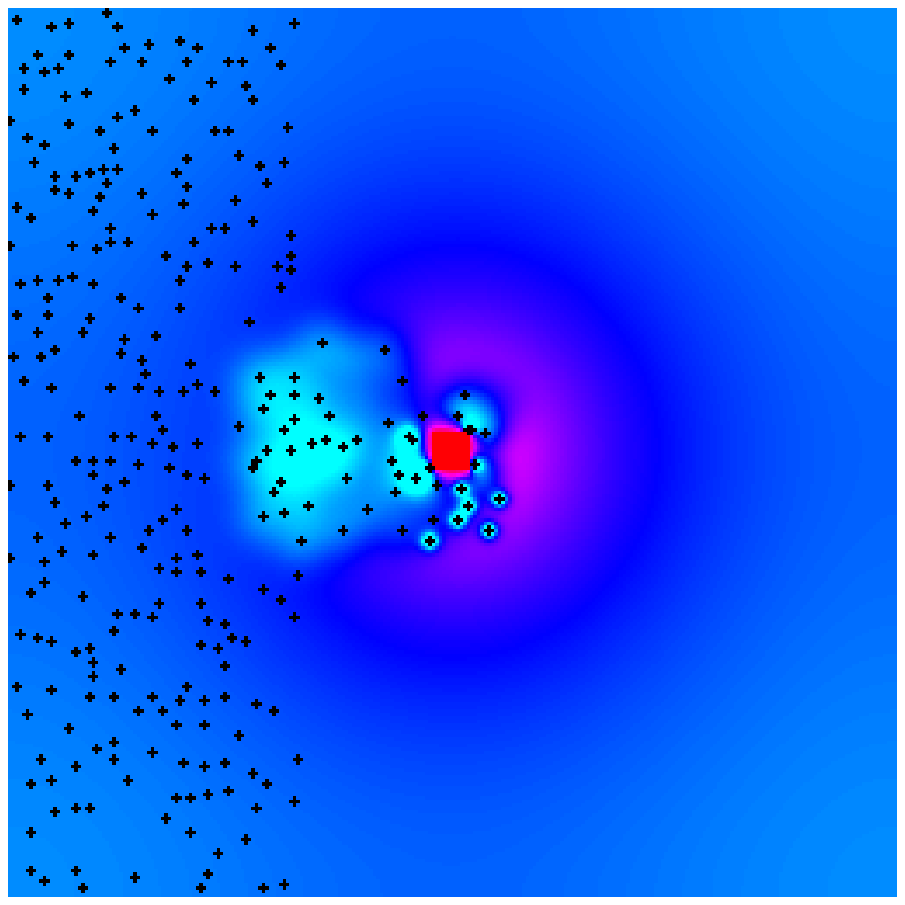}
    \includegraphics[scale=0.33]{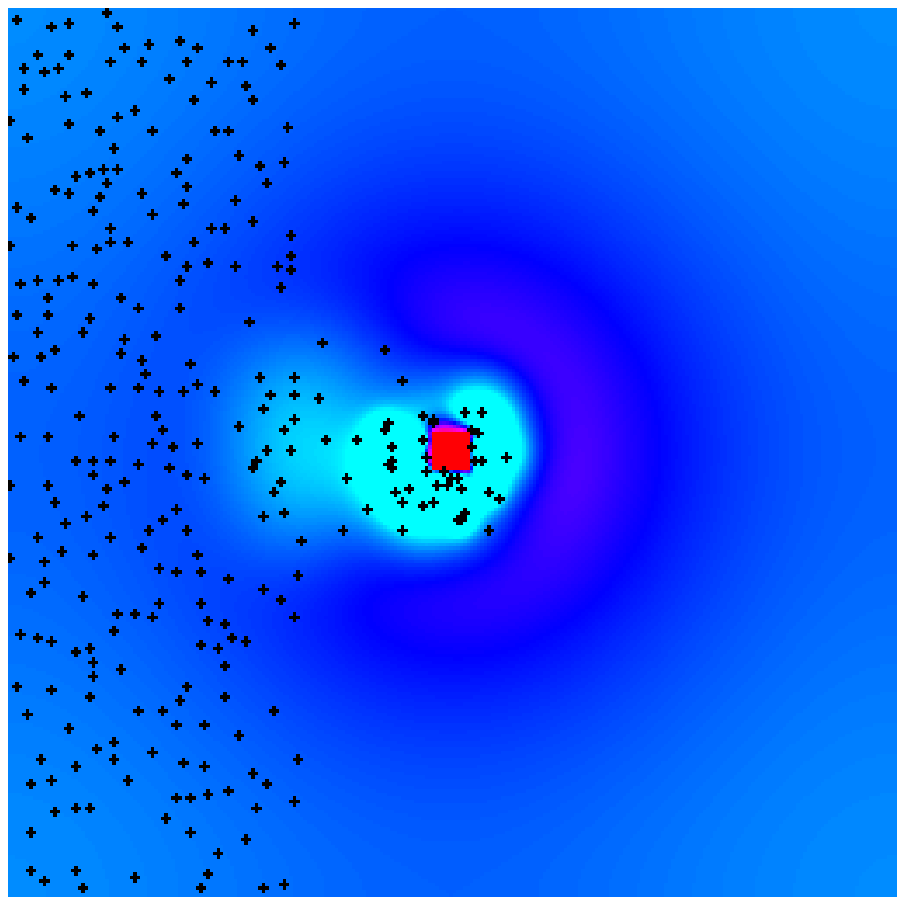}
    \includegraphics[scale=0.33]{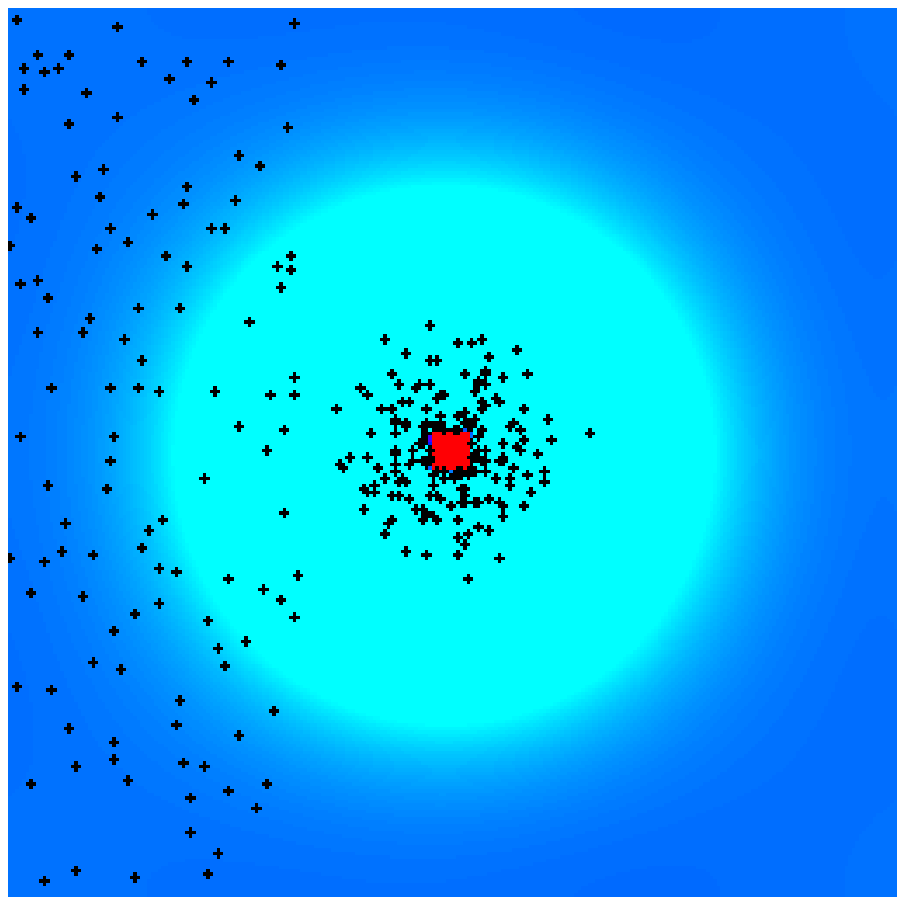}
    \includegraphics[scale=0.33]{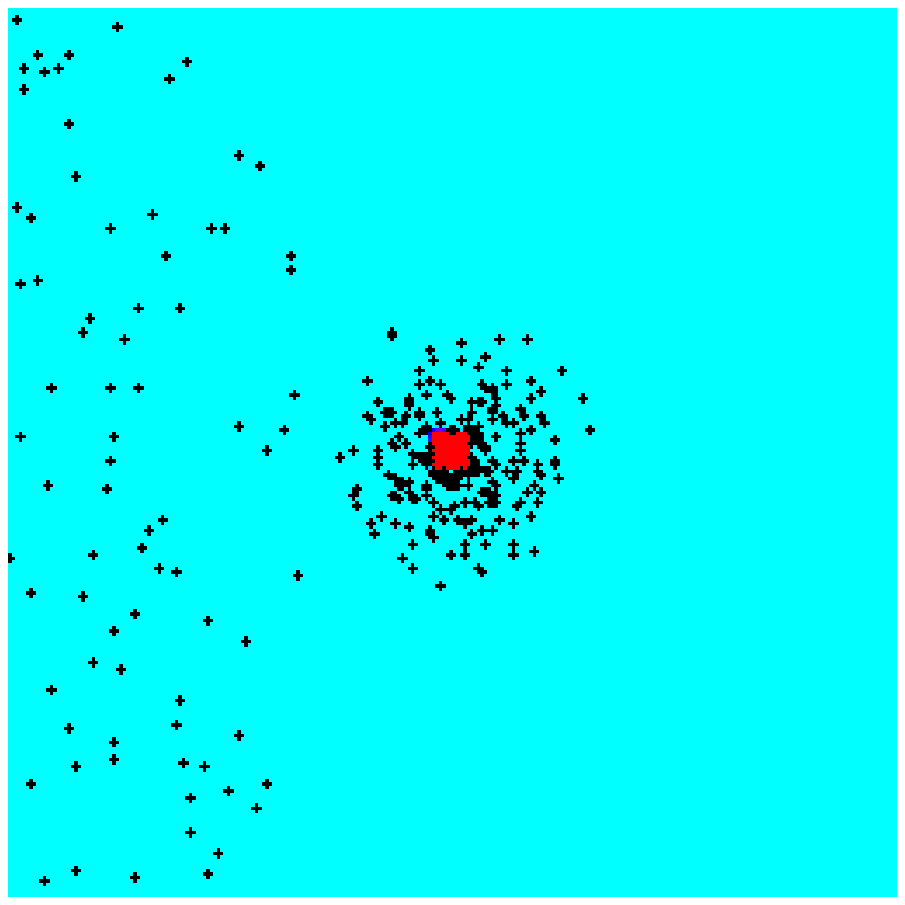}  \\
    $t=50668$ \hspace{1.3cm}   $t=50800$ \hspace{1.3cm} $t=51252$ \hspace{1.3cm} $t=60000$ \hspace{1.3cm} $t=80000$ \\
    (c) \\
    \includegraphics[scale=0.33]{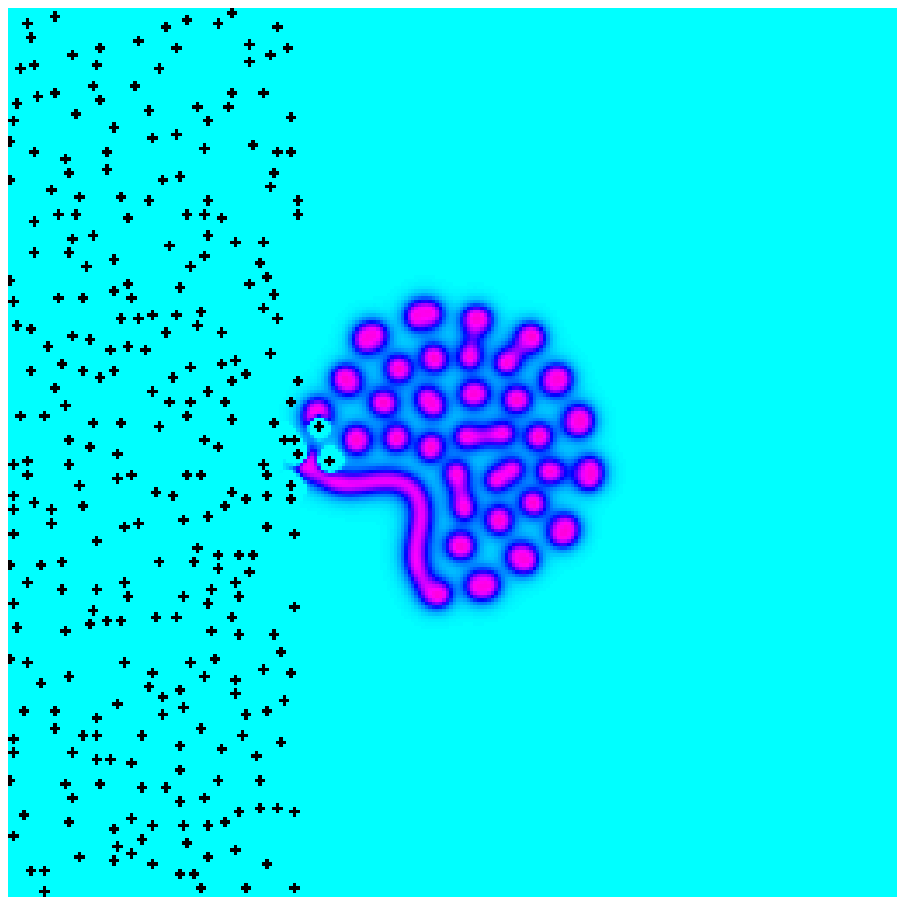}
    \includegraphics[scale=0.33]{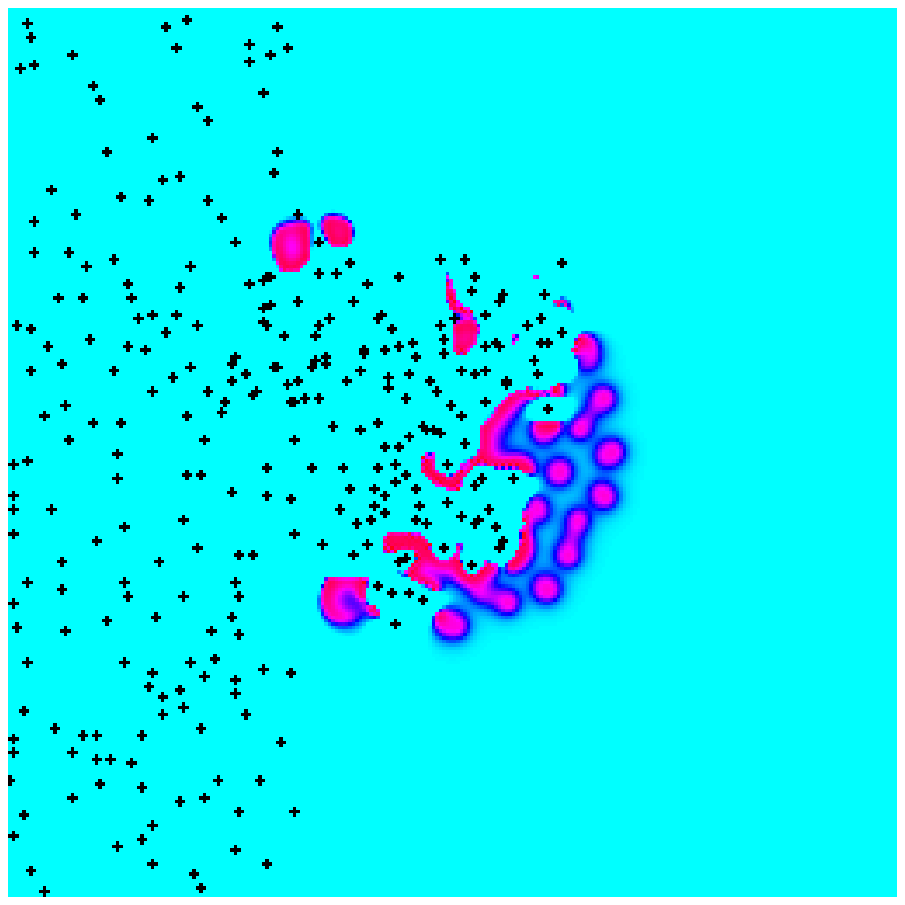}
    \includegraphics[scale=0.33]{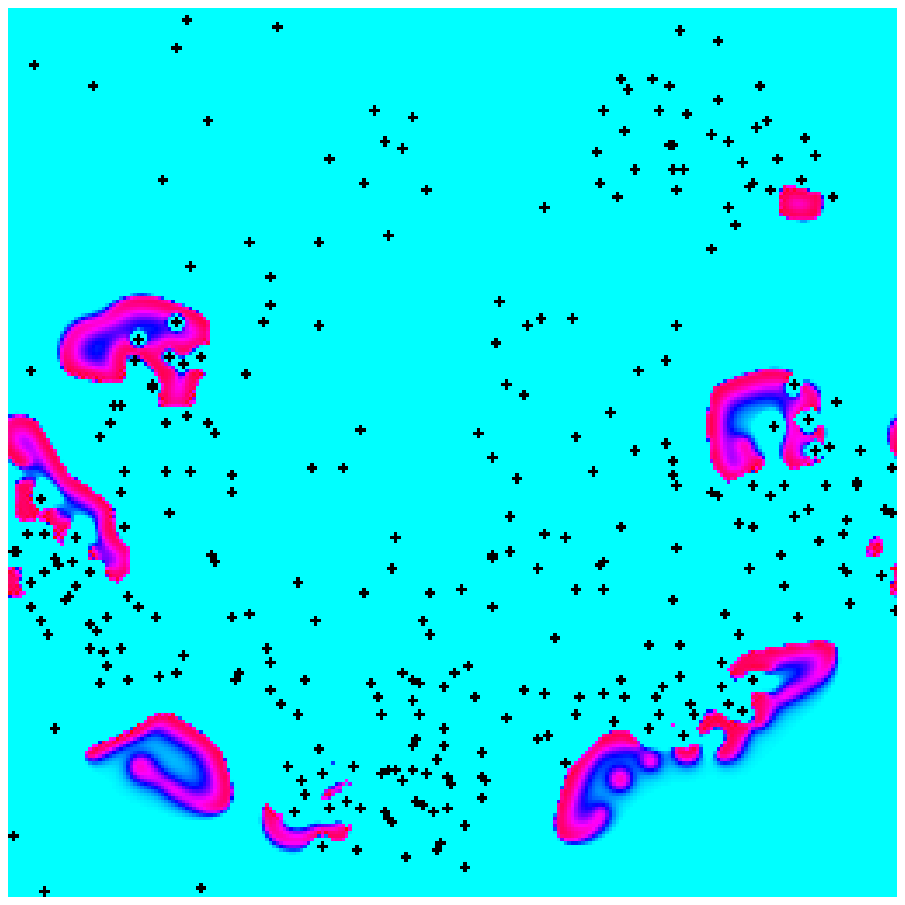}
    \includegraphics[scale=0.33]{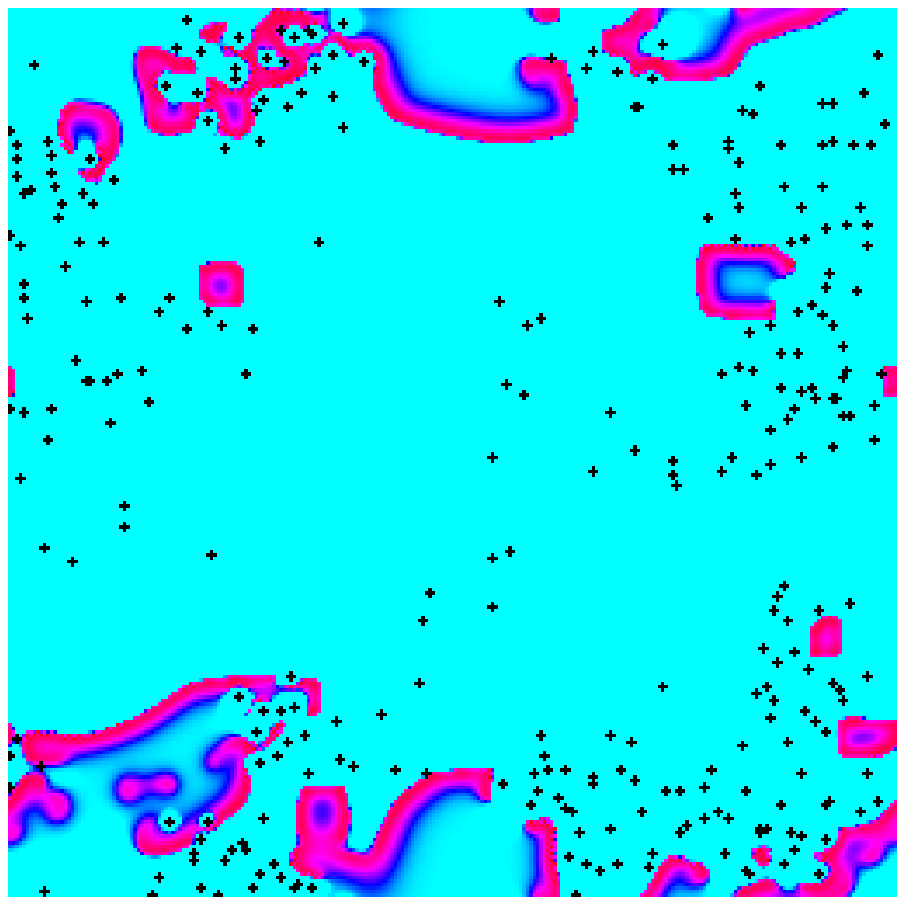}
    \includegraphics[scale=0.33]{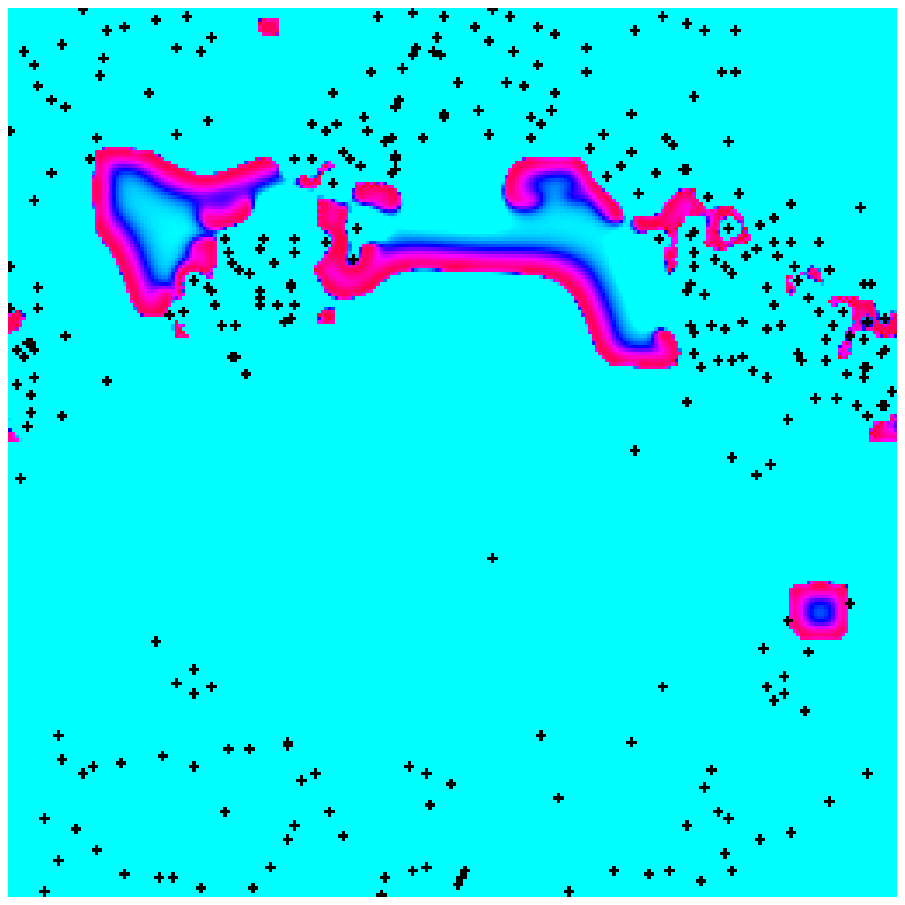}  \\
    $t=3940$ \hspace{1.4cm}   $t=4840$ \hspace{1.4cm} $t=6840$ \hspace{1.4cm} $t=7840$ \hspace{1.4cm} $t=8840$ \\
    (d) \\
    \includegraphics[scale=0.33]{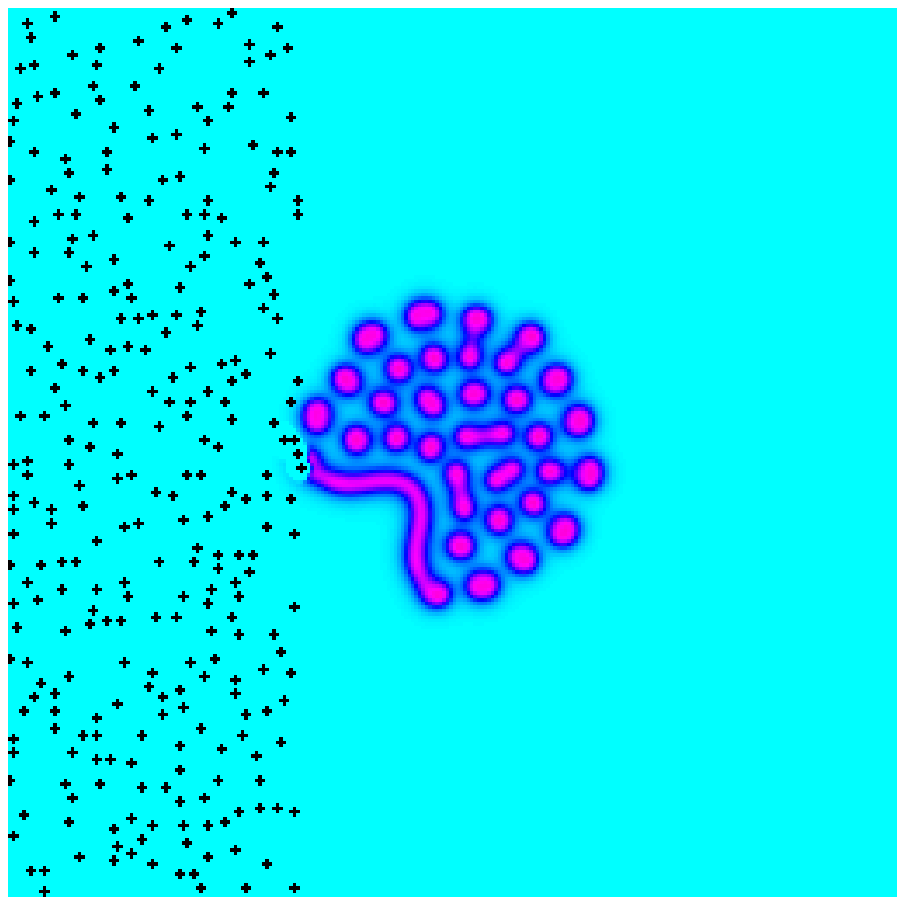}
    \includegraphics[scale=0.33]{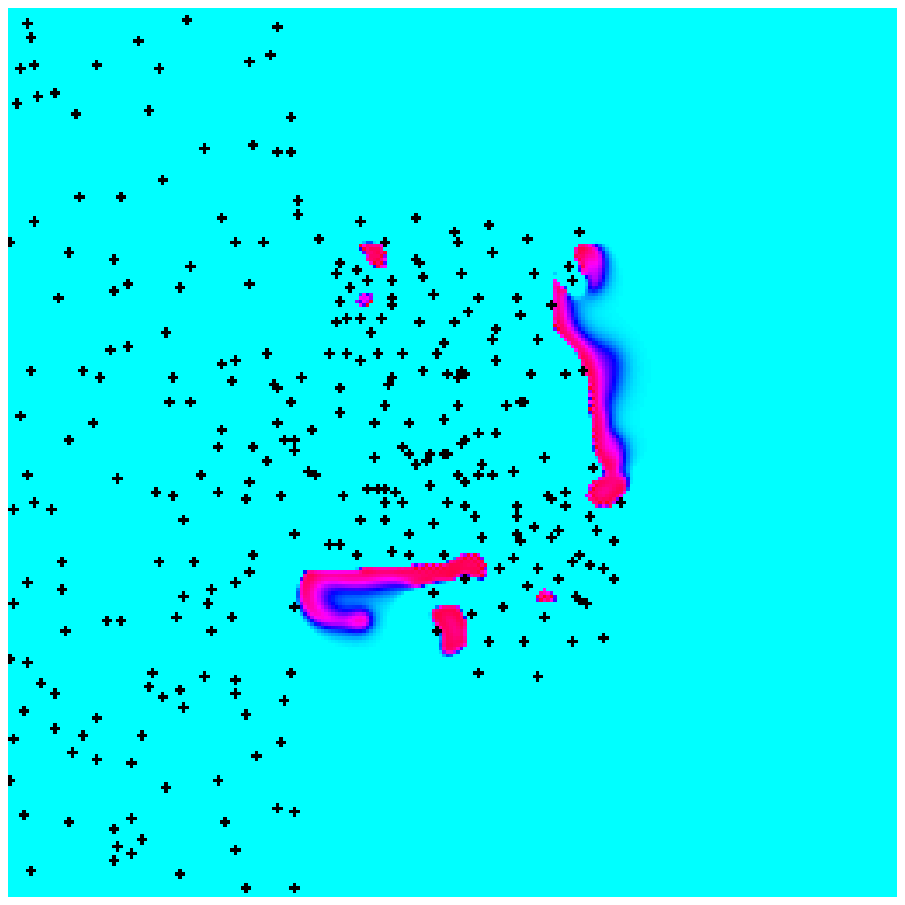}
    \includegraphics[scale=0.33]{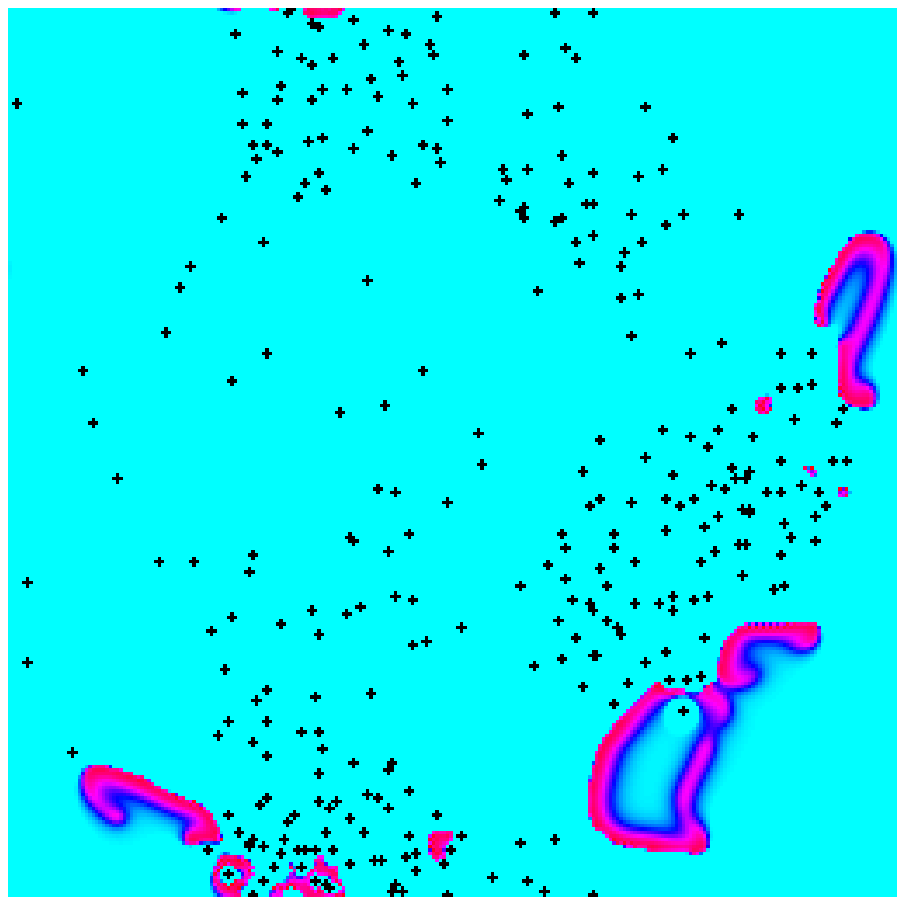}
    \includegraphics[scale=0.33]{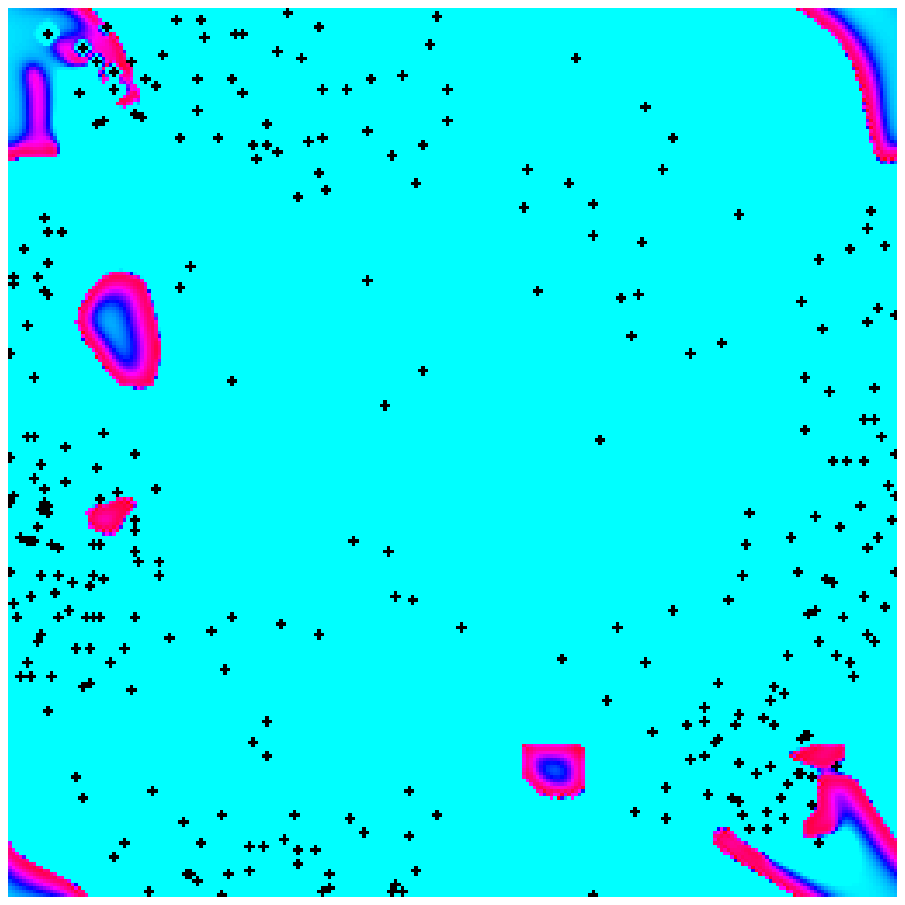}
    \includegraphics[scale=0.33]{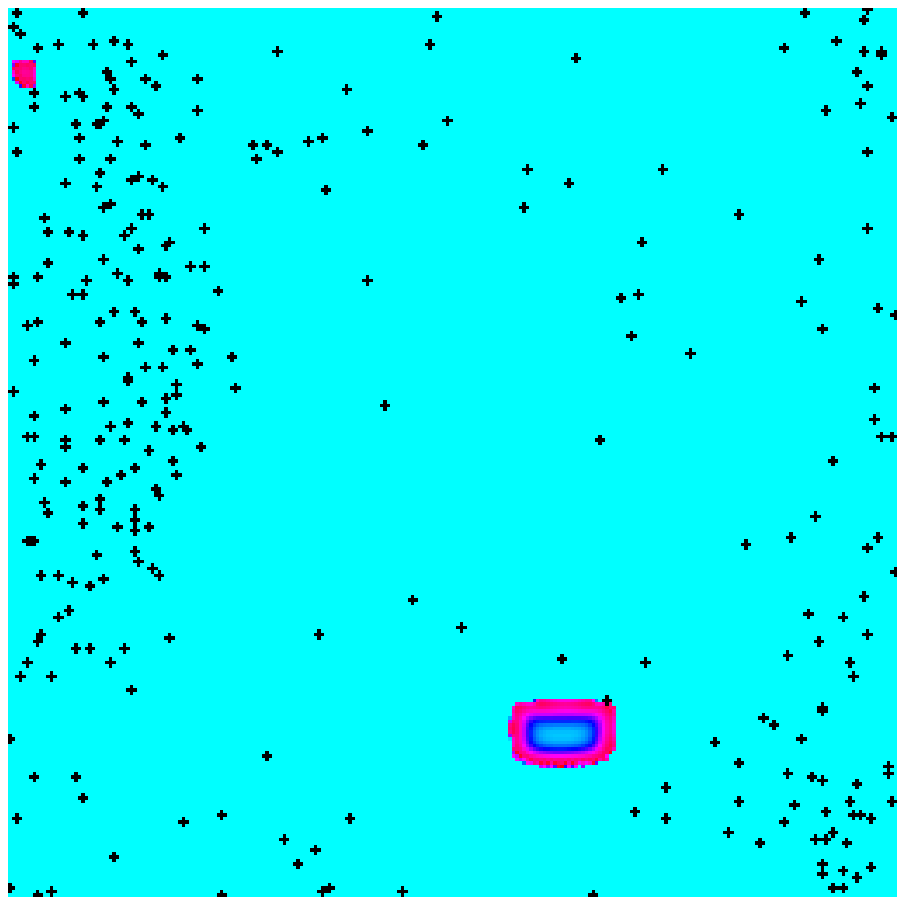}  \\
    $t=3940$ \hspace{1.4cm}   $t=4840$ \hspace{1.4cm} $t=6840$ \hspace{1.4cm} $t=7840$ \hspace{1.4cm} $t=8840$ \\
    \caption{Snapshots of the pattern for the four cases: (a) Fick
    diffusion and ER network, (b) Fick diffusion and BA network, (c)
    Gray-Scott reaction-diffusion and ER network and (d) Gray-Scott
    reaction-diffusion and BA network. Red (on-line version)
    represents maximum disease intensity and cyan (on-line version) no
    disease. The node are represented by the black dots. Both networks
    have $300$ nodes and $\langle k\rangle \approx 4$.}
  \label{fig:03}
  \end{center}
\end{figure*}

Figure~\ref{fig:03} presents snapshots of the evolution of the
disease for four cases assuming the \textit{one-source} configuration. A set of movies with all the configurations discussed in this paper can be viewed at \mbox{\textit{http://cyvision.ifsc.usp.br/$\sim$luisrocha/paper/}}. The
first two rows show the evolution of the Fick diffusion controlled by (a) ER and (b) BA defensive networks, both with $300$ node and $\langle k\rangle \approx 4$. In this case, the activation of the
first node only took place after a relatively long period of time.
More precisely, the first activation (not shown in the figures)
occurred about $8200$ time units before the first snapshot in the ER
case and about $3600$ time units in the BA case, but they had little
effect on disease control. However, these early activations were
important because they brought some node closer to the disease
source. As soon as some node were close enough to the source, they
were activated, triggering a chain activation effect (first snapshot in fig.~\ref{fig:03}-a). The latter effect occurred because some of the activated node fell at positions where the disease had
already overcame the threshold (second snapshot in
fig.~\ref{fig:03}-a). Hence, the spatially activation of the node
resulted on requests to their own neigbors, and so on. Observe the
\textit{hub} activation in the BA case (first snapshot in
fig.~\ref{fig:03}-b). In this case, many activated
node were requested at once and, consequently, some of them fell
very near the source. As a result, their own neigbors (\textit{i.e.} the
neigbors of the neigbors of the hub) were activated, consequently
populating the area around the source and enclosing it with a
considerable amount of antidote (second snapshot in
fig.~\ref{fig:03}-b). The ER network node took three times longer
to achieve the control of the source (third snapshot in
fig.~\ref{fig:03}-a), \textit{i.e.} to encircle the source. After this stage
of the chain reaction, the source became enclosed and the node kept
on changing their states. Each new spatial activation redistributed
the helpers around the source and even requested node which had
never been activated before. The latter effect, \textit{i.e.} the
activation of the \textit{hubs}, implied in the fastest decrease in
the total quantity of the disease considering the BA network
(fourth snapshot in fig.~\ref{fig:03}-b). At the last considered snapshot, the mesh was
found to be more free of disease in the BA case, while a substantially
more infected configuration was observed in the ER cases (fifth snapshot in
fig.~\ref{fig:03}-a). After very long times, the in node ended to converge around
the source.

The chosen configuration of the Gray-Scott reaction-diffusion
($f=0.04$ and $k=0.064$) generated non-static patterns whose spots and
stripes tended to quickly reach the complex network. Figures~\ref{fig:03}-c
and~\ref{fig:03}-d, represent the reaction-diffusion evolution
constrained by the ER and BA defensive networks, respectively, with
$300$ node and $\langle k\rangle \approx 4$ each. After the first
node activation (first snapshot in fig.~\ref{fig:03}-c,d), a chain
reaction was triggered as in the Fick diffusion model. The node were
activated from the center to the boundary of each case (second
snapshot in fig.~\ref{fig:03}-c,d), a natural consequence of the
dynamics rules. Once again, the \textit{hub}-based characteristic of
the BA network resulted on massive attack against the disease. This
type of attack can be identified by the great amount of eliminated
disease in the reaction-diffusion constrained by the BA network
(second snapshot in fig.~\ref{fig:03}-d) in contrast to the ER
network (second snapshot in fig.~\ref{fig:03}-c). Because of the
finite-size and sparse connectivity of both types of network, not
enough neigbors nodes were requested, allowing leakage and subsequent
relapsing of the disease. Due to the antidote liberation, the disease
grew in the direction contrary to where the node were placed. Even
the small disease sources of the BA case produced much infection in
the mesh after about $3000$ time units from the first activation (third 
snapshot in fig.~\ref{fig:03}-d). However, the non-massive attacks of
ER network nodes resulted on more isolated patterns and in faster increase
of the disease quantity (third snapshot in fig.~\ref{fig:03}-c). After the interval of increase (fourth 
snapshot in fig.~\ref{fig:03}-c,d), the node retook control,
eliminating many isolated patterns (fifth snapshot in
fig.~\ref{fig:03}-c,d). While the node were eliminating many
isolated patterns (fifth snapshot in fig.~\ref{fig:03}-c), a uniform spatial
node distribution emerged in the mesh. Conversely, in the presence of
few infected areas, the nodes joined efforts to eliminate them and
concentrated themselves on the highest infected regions of the mesh
(fifth snapshot in fig.~\ref{fig:03}-d). After the complete
elimination of the disease, the node remained on their last
respective positions. Observe that the original
pattern was modified at the places where the antidote acted, specially
near the activated node.

The ability of the defensive network to control and stop the disease was verified to be directly related to the number of node and to the connectivity of the network. We expected that with more node being
activated, they would more readily gather control and completely
eliminate the disease spreading. A larger and completely connected
network would activate all the neigbors at once and hence fully
populate the mesh. Consequently, the disease would fade down quickly
until complete elimination. Such a network would imply high
maintenance costs if adopted by natural (or artificial) systems. In
fact, it is often mandatory to achieve maximum efficiency by using the minimum
amount of energy. In practice, many of the networks which have been
investigated in complex networks research are characterized by low
connectivity among their node~\cite{Newman_review, Dorog_review,
BaraAlbert_review}. Therefore, it is interesting to investigate the
efficiency of ER and BA networks with small number of node
(relatively to the mesh size) and low connectivity, as in many natural
and artificial real systems.

\begin{figure}[ht]
  \begin{center}
  \includegraphics[scale=0.2]{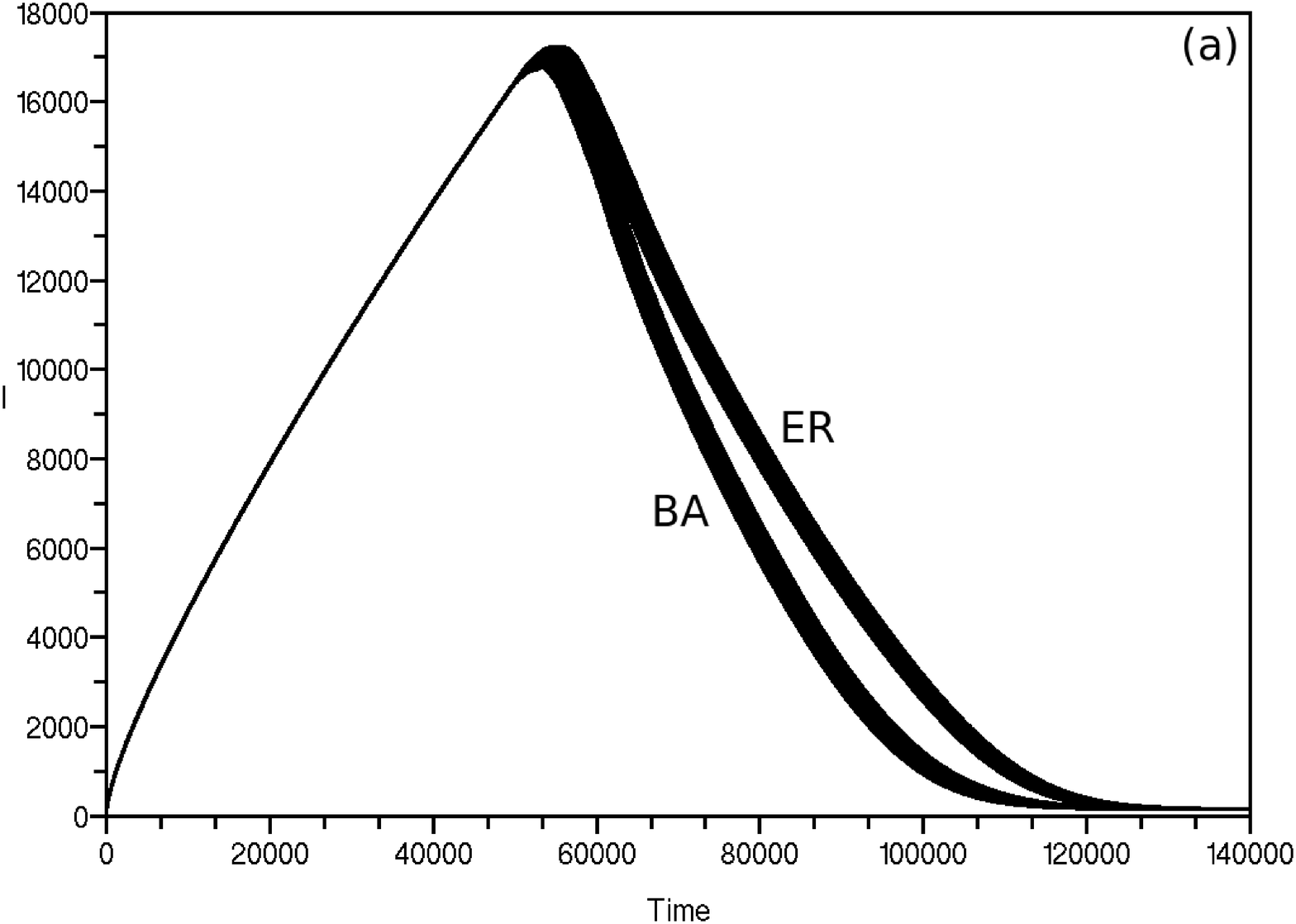}
  \includegraphics[scale=0.2]{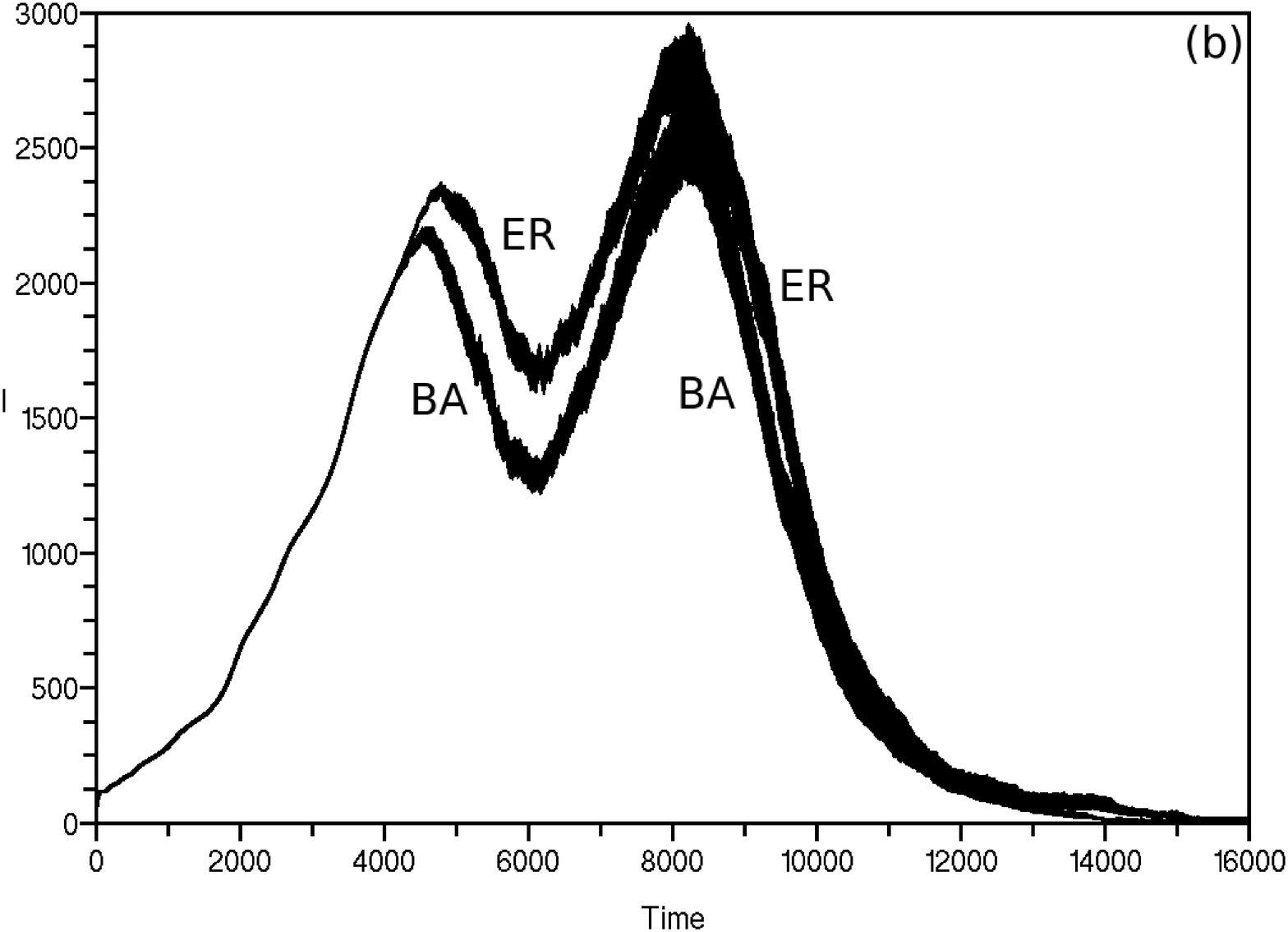}
  \caption{The amount of disease $I$ in the mesh ($y-axis$) at time
  ($x-axis$). (a) Fick diffusion model and (b) Gray-Scott model. \textit{One
  source}, $N=300$ and $\langle k\rangle \approx 4$ configuration. The
  standard deviation is (a) one fifth and (b) one tenth of the real
  value. } \label{fig:04}
  \end{center}
\end{figure}

\begin{figure}[ht]
  \begin{center}
  \includegraphics[scale=0.2]{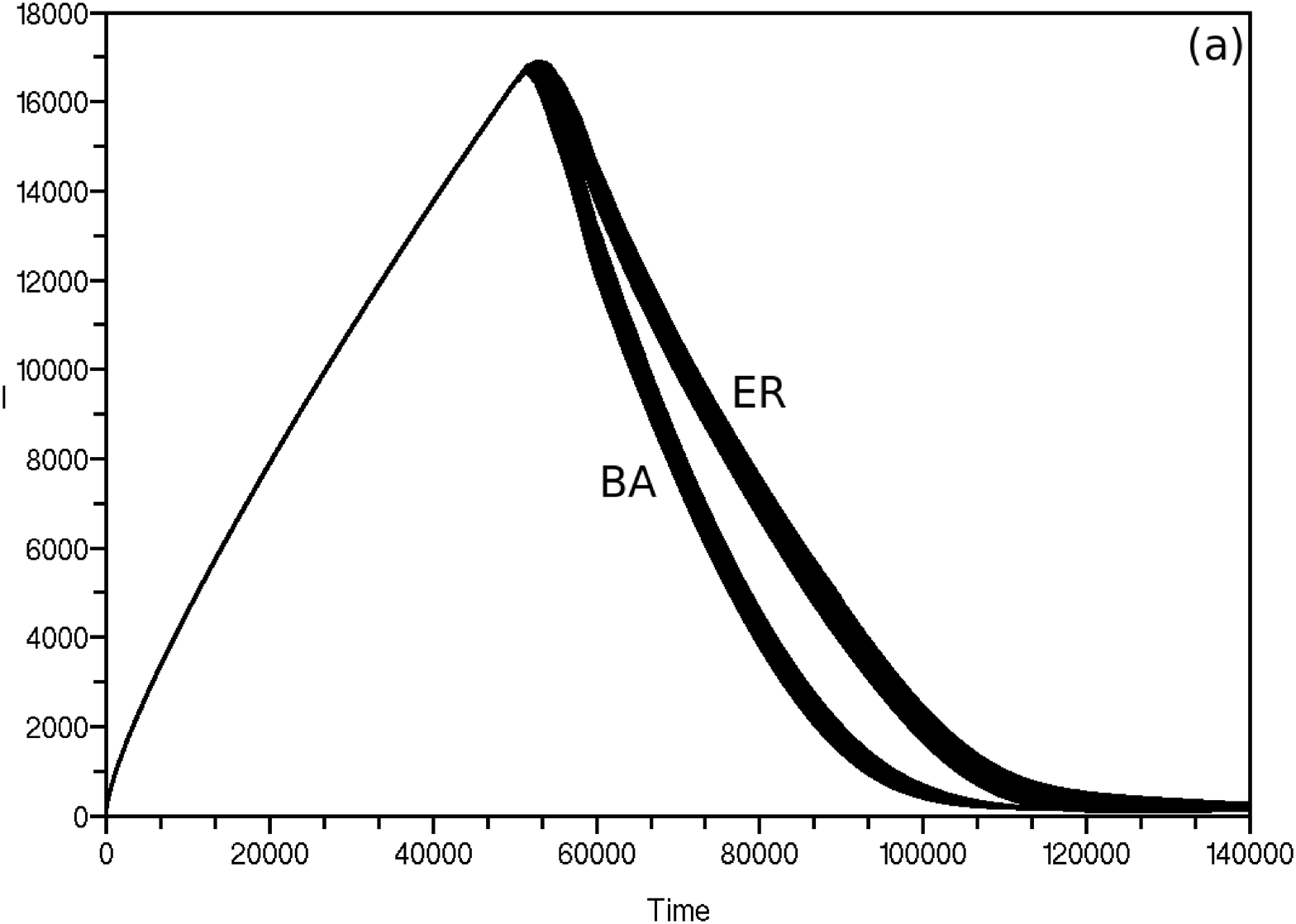}
  \includegraphics[scale=0.2]{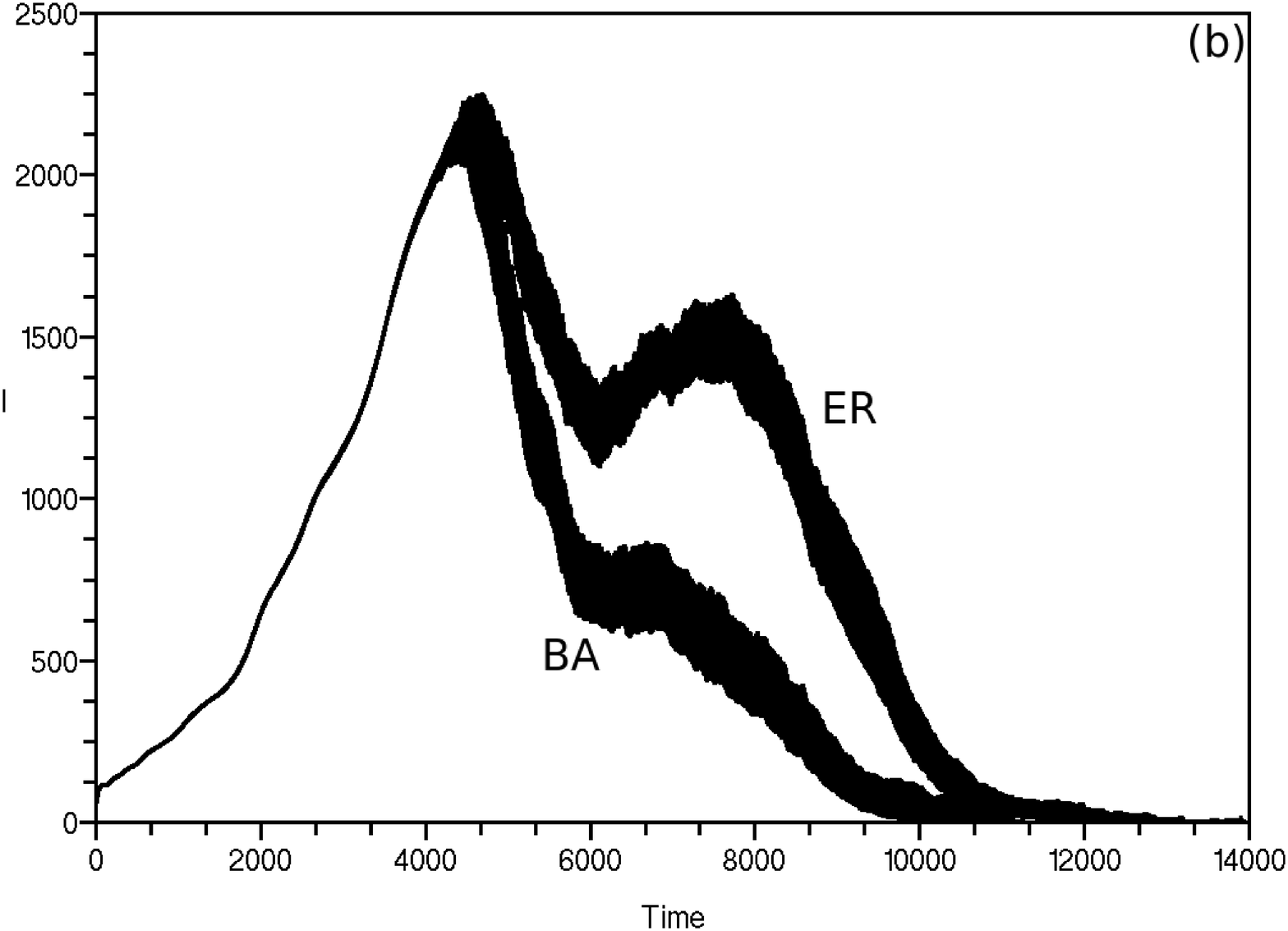}
  \caption{The amount of disease $I$ in the mesh ($y-axis$) in terms of time
  ($x-axis$). (a) Fick diffusion model and (b) Gray-Scott model. \textit{One
  source}, $N=500$ and $\langle k\rangle \approx 4$ configuration. The
  standard deviations in this figure corresponds to
  one fifth of their real values. } \label{fig:05}
  \end{center}
\end{figure}

\begin{figure}[ht]
  \begin{center}  
  \includegraphics[scale=0.2]{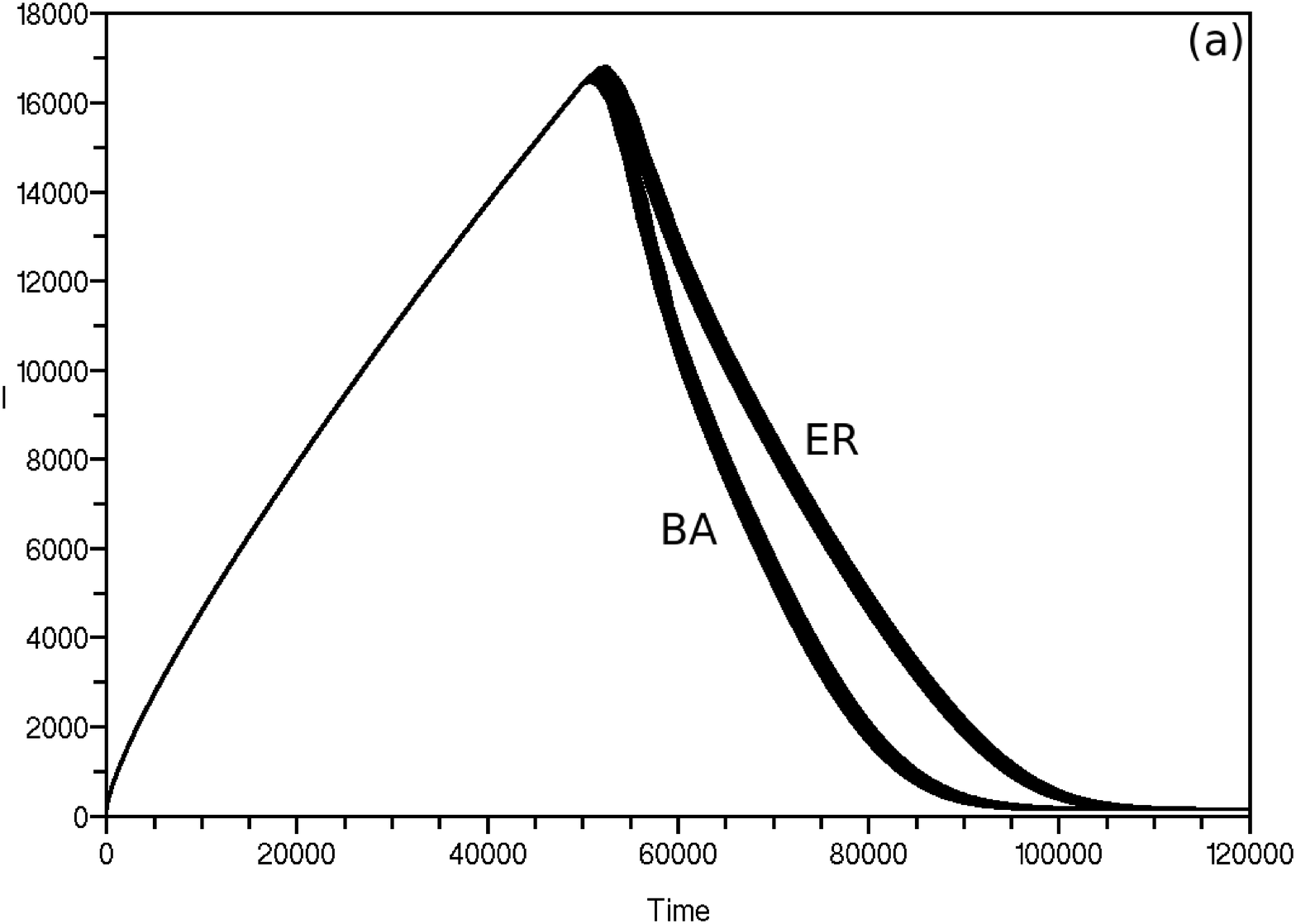}
  \includegraphics[scale=0.2]{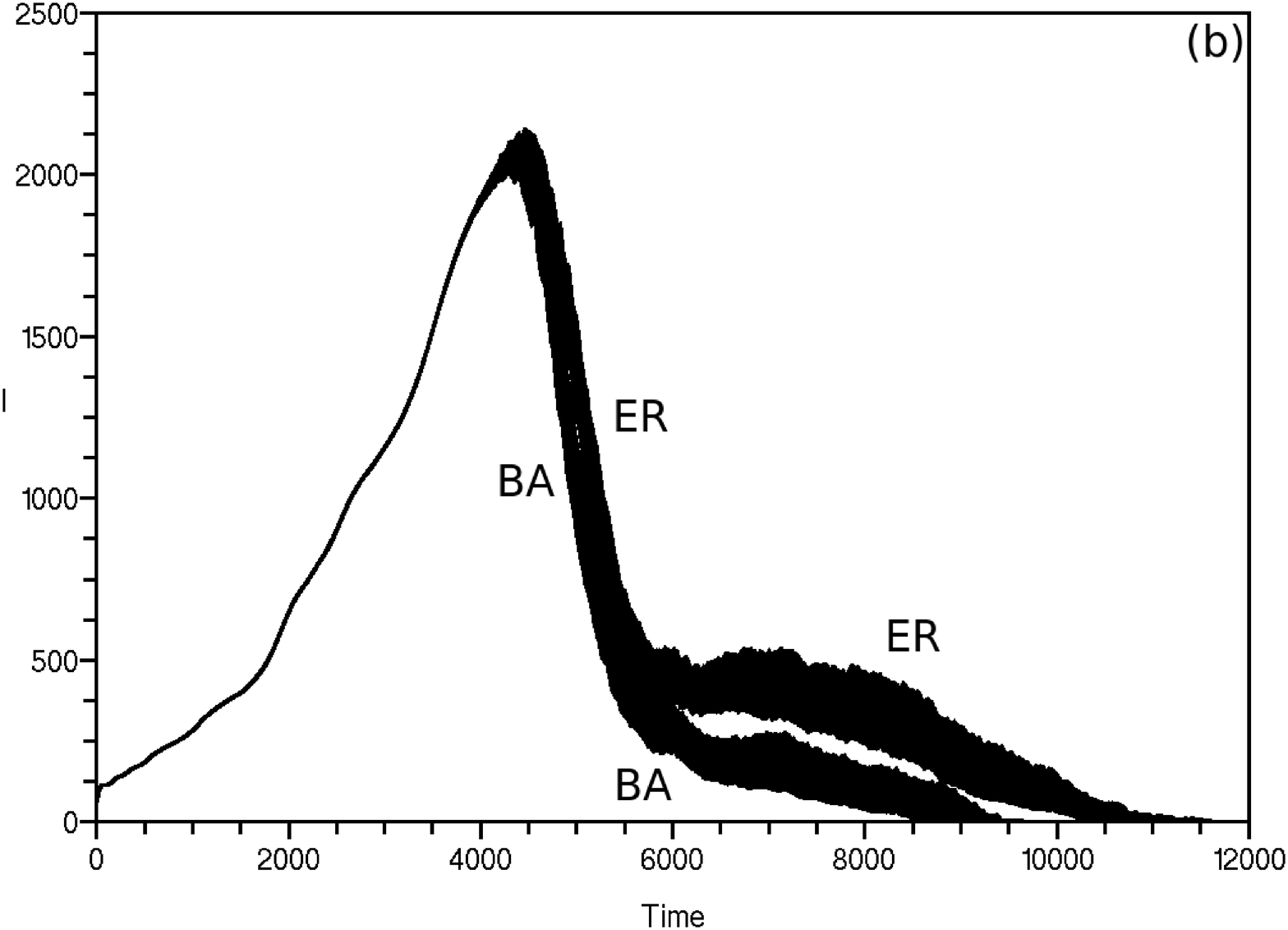}
  \caption{The amount of disease $I$ in the mesh ($y-axis$) at time
  ($x-axis$). (a) Fick diffusion model and (b) Gray-Scott model. \textit{One
  source}, $N=500$ and $\langle k\rangle \approx 6$ configuration. The
  shown standard deviation is one fifth of its real value. } \label{fig:06}
  \end{center}
\end{figure}

Figure~\ref{fig:04}-a compares the evolution of the amount of
disease $I$ for the Fick diffusion model using the \textit{one-source}
configuration. A total of $100$ realizations was considered for
each parameter configuration. The quantity $I$ had a nearly constant
growth rate up to a maximum, when the first node were activated.
These node triggered a chain reaction, but on the average both
networks had similar efficiency in controlling the diffusion in the
beginning, \textit{i.e.}, until about $60000$ time units. The time
spent to enclose the source was relatively short and, on average, no
difference could be observed between both types of networks. The
importance of \textit{hub}-activation, implying liberation of more
antidote, showed up after $60000$ time units, when the diffusion
constrained by the BA network clearly decreased faster than the
diffusion observed for the ER network. By comparing
figures~\ref{fig:04}-a,~\ref{fig:05}-a and~\ref{fig:06}-a, it is
clear, for both types of networks, that the diffusion dropped down
and reached minimum levels faster when the number of defensive node
and the connectivity of them were increased. However, BA 
network nodes continued to be more effective against the disease than the ER
network nodes. Interestingly, the minimal level of diffusion was
reached at nearly the same time in both types of networks in the
first configuration ($N=300$ and $\langle k\rangle \approx 4$) and
about $20000$ time units earlier in the BA case than in the ER case
for the other two configurations ($N=500$, $\langle k\rangle \approx
4$ and $\langle k\rangle \approx 6$), a consequence of the increased amount of antidote liberated in the first stages of the defense.

The non-uniform patterns generated by the Gray-Scott reaction-diffusion implied richer dynamics (fig.~\ref{fig:03}).
Starting from the initial source, non-localized patterns emerged over time, creating fast moving spots and stripes, so that the nodes had to actively move through the regular network in order to eliminate the disease. The
amount of disease increased in nearly quadratic fashion with time up
to a maximum when the first node were activated. Depending on the
connectivity of the defensive network, different evolutions were
clearly obtained after the first activation. The reaction-diffusion
constrained by networks with $\langle k\rangle \approx 4$ (see 
fig.~\ref{fig:04}-b and fig.~\ref{fig:05}-b), resulted on three stages: (i)
a decrease down to a minimum level, (ii) a relapse up to a local
maximum level and (iii) resumption of the decrease until the disease
is eliminated. On the other side, the network with $\langle k\rangle
\approx 6$ (see fig.~\ref{fig:06}-b) exhibited two stages: (i) fast and
(ii) slow elimination of the disease. Observe that this phenomenon
is not only due to the connectivity, but also depends on the number
of nodes: a higher quantity of simultaneously activated nodes
resulted in more antidote and, consequently, reduction of the
disease (\textit{i.e.} $I$).

The first stage of defense (between $4000$ and $6000$ time units),
was a consequence of \textit{hierarchical neighbors}
activation~\cite{CostaRocha_generalized}. Once the disease had
considerably diffused along the space, every new request contributed
to the distribution of more activated nodes radially to the boundary
of the disease. Obviously, some nodes fell on positions without
disease.  It was also possible to have some nodes requested by their
own requested neigbors. Because of the finite-size of the network,
on the average the hierarchical number of neighbors had a peak
$n_{max}$ whose value depends on the number of nodes and on the
connectivity of the network~\cite{CostaNascimento_hierarchy}.
The presence of \textit{hubs} implied that $n_{max}$ is
reached faster (in terms of \textit{hierarchical levels}) in the BA than in
the ER network. In other words, BA node activate more neigbors at
once than the ER node in the first \textit{hierarchical levels}. Therefore, in this stage
the disease decreased faster in the BA than the ER case as shown
in Figures~\ref{fig:04}-b,~\ref{fig:05}-b and~\ref{fig:06}-b. As
expected, more node and higher connectivity implied on more effective 
decrease in the disease intensity.

The second stage of defense (between $6000$ and about $8000$ time
units) was characterized by leakage of disease from the first massive attack
(\textit{i.e.}, chain reaction). The requested neigbors, in the
first stage, were not enough to control the disease, \textit{i.e.} although
they broke the pattern, some isolated regions of disease
concentration remained which resumed progression.  ER networks
tended to engage less nodes than BA networks, allowing the creation
of a larger number of isolated patterns. The latter effect implied
more competition for node, postponing the control of the disease.
Figures~\ref{fig:04}-b and ~\ref{fig:05}-b show that as the number
of network nodes was increased, the relapse peak tended to diminish,
\textit{i.e.}, more node resulted in more effective control of
the disease. The relapse peak depended considerably of the height of
the disease intensity $I$ at the turning point~\footnote{The turning
point corresponds to the abscissae of the relative minimum of the
disease intensity $I$, which tended to occur nearly after $6000$
steps.}, \textit{i.e.}. More distributed patterns implied in more
intense relapse and increased difficult of respective control. The
node had to swap their places constantly, following the requests
which depended on the connectivity of the defensive network and not
on the node distance in the regular network. Consequently, these
movements of node resulted on vacancies in the regular network,
which allowed the local development of disease. Observe that the
effective elimination of disease by the latter network
configuration, \textit{i.e.}, $N=500$ and $\langle k\rangle \approx
6$ (fig.~\ref{fig:06}-b), resulted in few remaining sources.
Consequently, the defensive network was capable to control the disease and
maintained a low level of disease prior to its complete elimination.

\begin{figure}[ht]
  \begin{center}
  \includegraphics[scale=0.2]{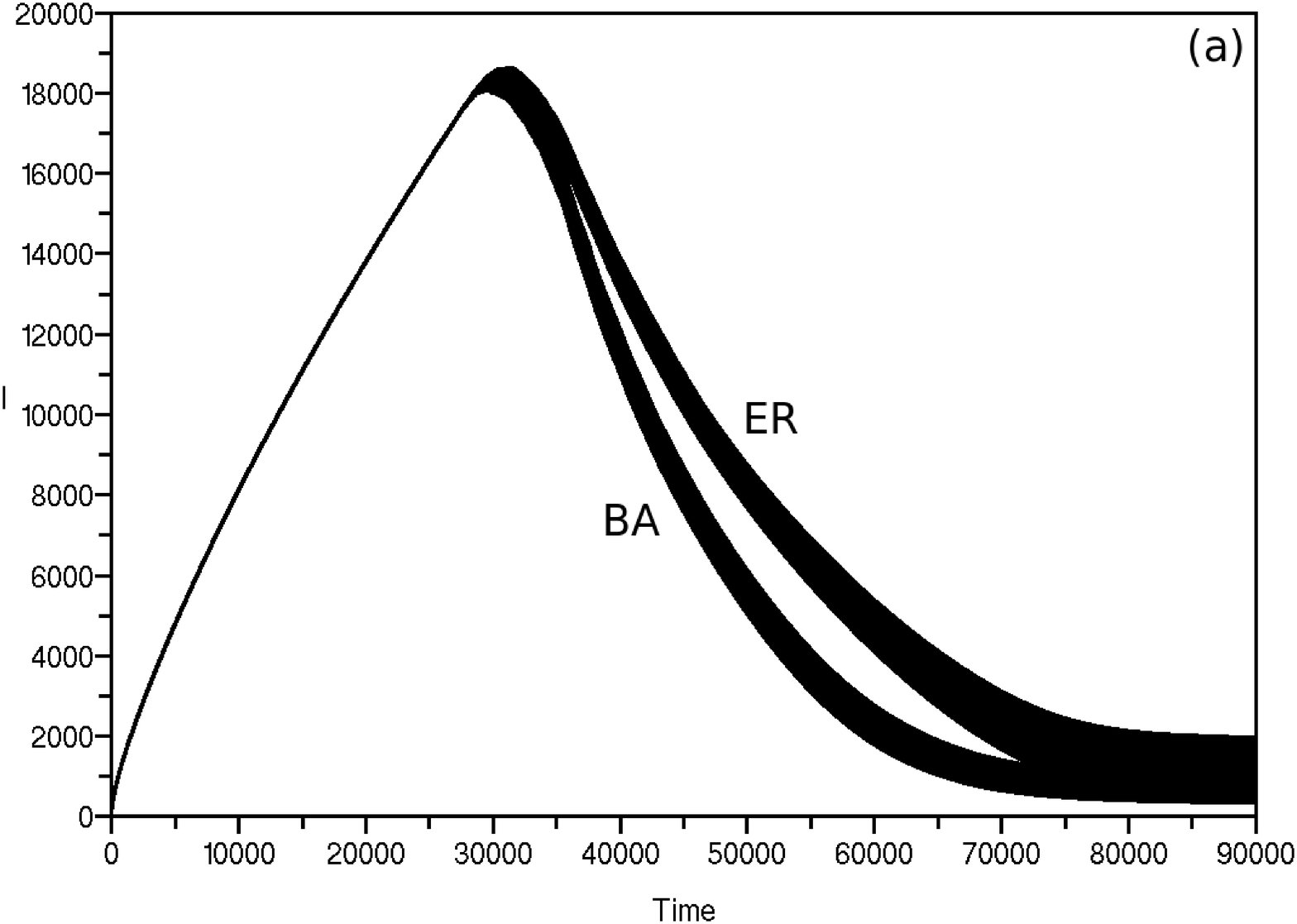}
  \includegraphics[scale=0.2]{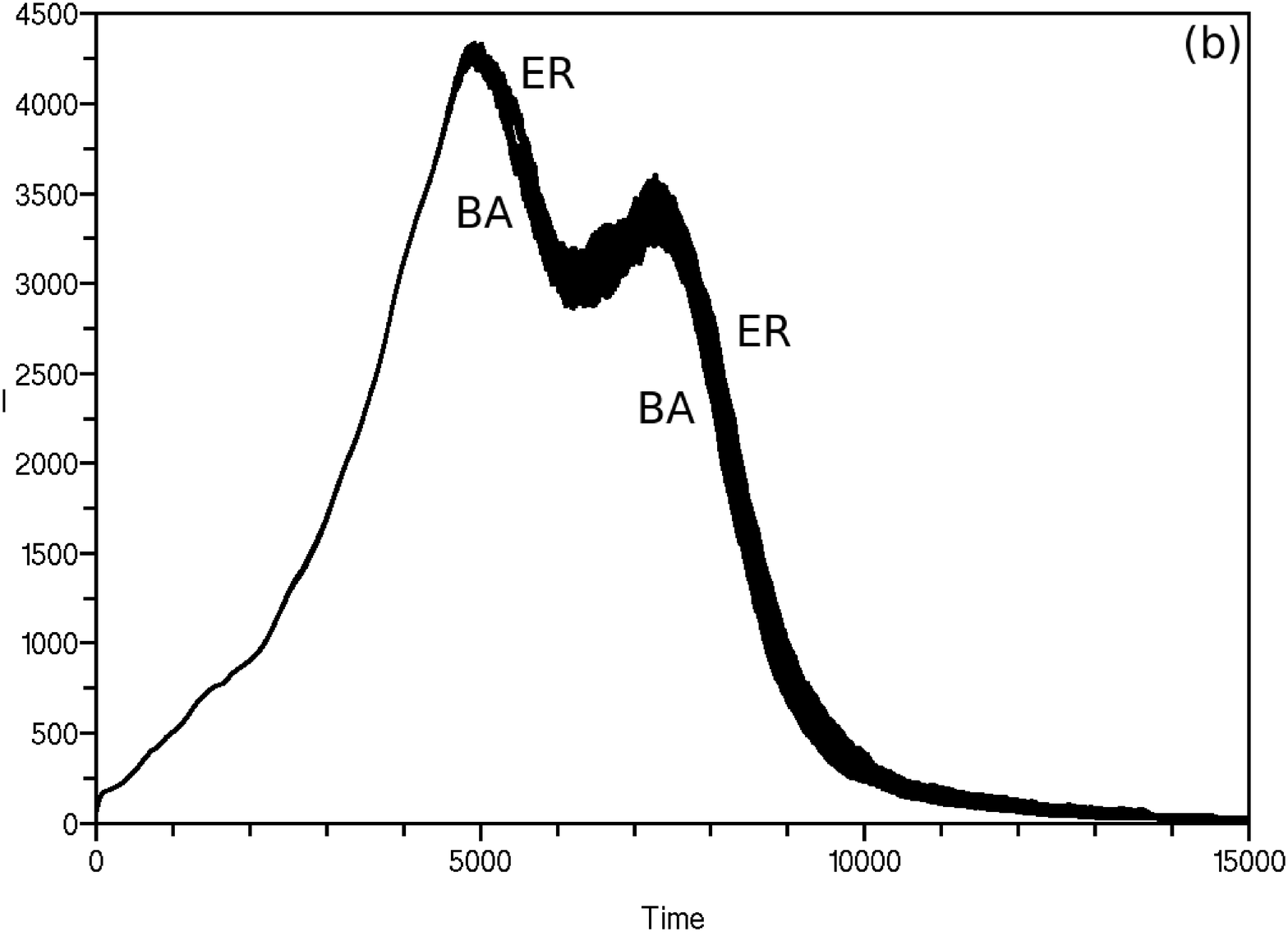}
  \caption{The amount of disease $I$ in the mesh ($y-axis$) along time
  ($x-axis$). (a) Fick diffusion model and (b) Gray-Scott model. \textit{Two
  sources}, $N=300$ and $\langle k\rangle \approx 4$ configuration. The
  standard deviation are shown at one fifth (a) and one tenth (b)
  of their real values.}  \label{fig:07}
  \end{center}
\end{figure}

\begin{figure}[ht]
  \begin{center}
  \includegraphics[scale=0.2]{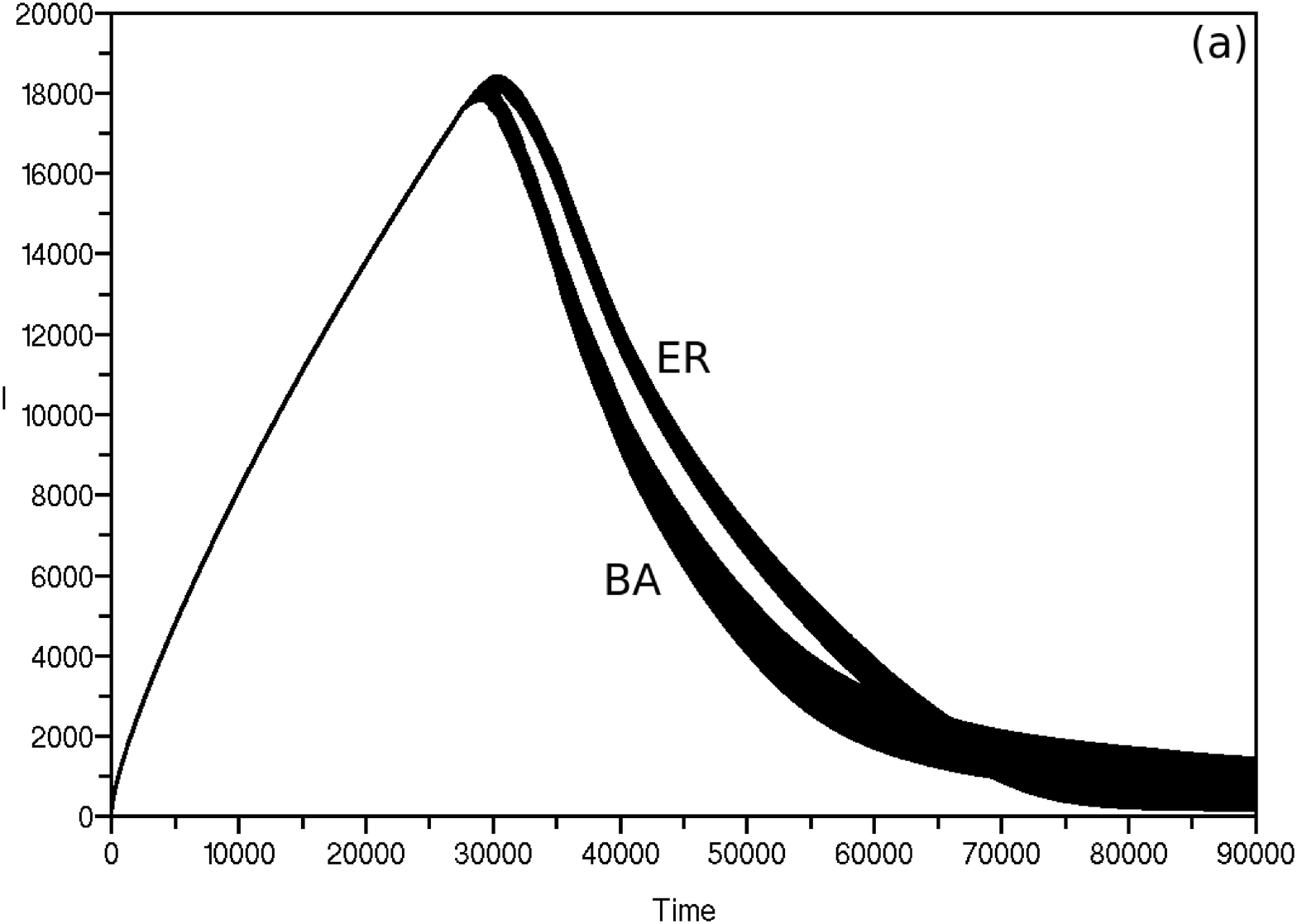}
  \includegraphics[scale=0.2]{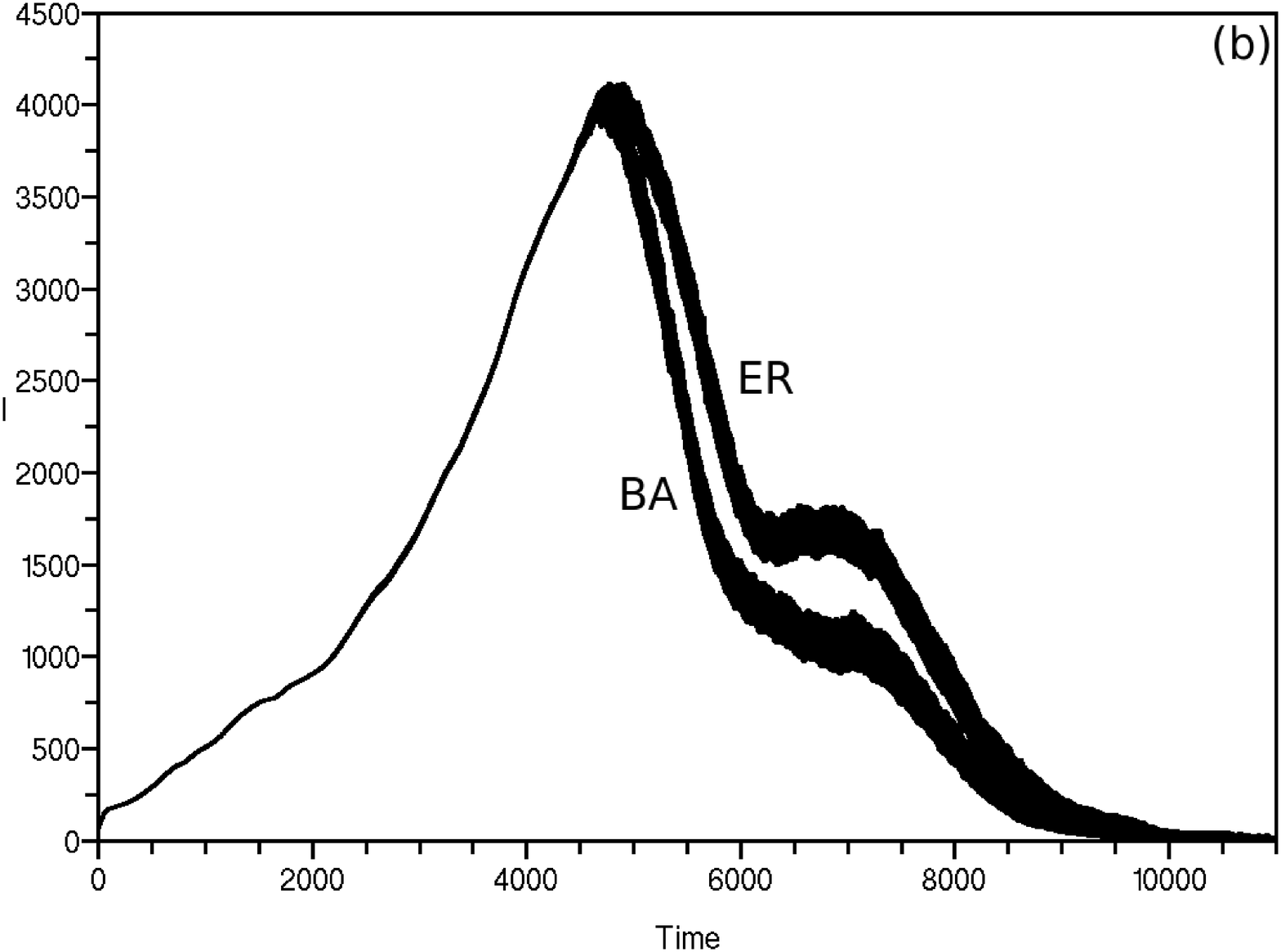}
  \caption{The amount of disease $I$ in the mesh ($y-axis$) along time
  ($x-axis$). (a) Fick diffusion model and (b) Gray-Scott model. \textit{Two
  sources}, $N=500$ and $\langle k\rangle \approx 4$ configuration. The
  standard deviation is shown at one fifth of its real value.}  \label{fig:08}
  \end{center}
\end{figure}

The third stage (between $8000$ time units and the complete disease
elimination, see fig.~\ref{fig:04}-b and fig.~\ref{fig:05}-b - absent
in the third network configuration shown in Figure~\ref{fig:06}-b) was a
consequence of the recovery of control by the activated nodes. Recall
that due to the initial conditions, much antidote was liberated in the
central area of the board in the first stage of defense. Naturally,
the disease grew faster in the antidote-free regions,
\textit{e.g.} opposite to the central area. However, much antidote
was also concentrated in other regions over time. This amount of
already liberated antidote contributed to slowed down the growing rate
of the disease.  The fact that the node had lesser disease to
eliminate contributed to faster elimination of the
infection. Interestingly, the long tail in the graphic in
Figures~\ref{fig:06}-b,~\ref{fig:07}-b and~\ref{fig:08}-b was a result of some small steady sources enclosed
by the antidote, these sources was not eliminated but could not grew too. Under this situation, an equilibrium was ultimately established where any growth of the disease was promptly eliminated by antidote being liberated by the surrounding node.  The latter behavior was also identified before the disease elimination in the third network configuration (fig.~\ref{fig:06}-b).

The competition for node played a fundamental role in the proposed
dynamics since help requests implied on depletion of node which
were previously activated.  If a neighbor $j$ was helping a node and
another node requested help from $j$, that node changes its position
with $50$ per cent of probability. Recall that node request as a
consequence of high activity in the regular network has priority
over solicitations by neighboring nodes.  As a consequence, only
regular network activated nodes did not change their positions while
at this state. Given the degree distribution of ER and BA networks,
we expected improvement in the ability of disease control to be
observed for the ER network. The more uniform distribution of
degrees in the former type of network resulted in a better
management of the distribution of node among many disease focuses.
Conversely, the request of many node by \textit{hubs} tended to
unbalance the number of node at each infected area.

\begin{figure}[ht]
  \begin{center}
  \includegraphics[scale=0.2]{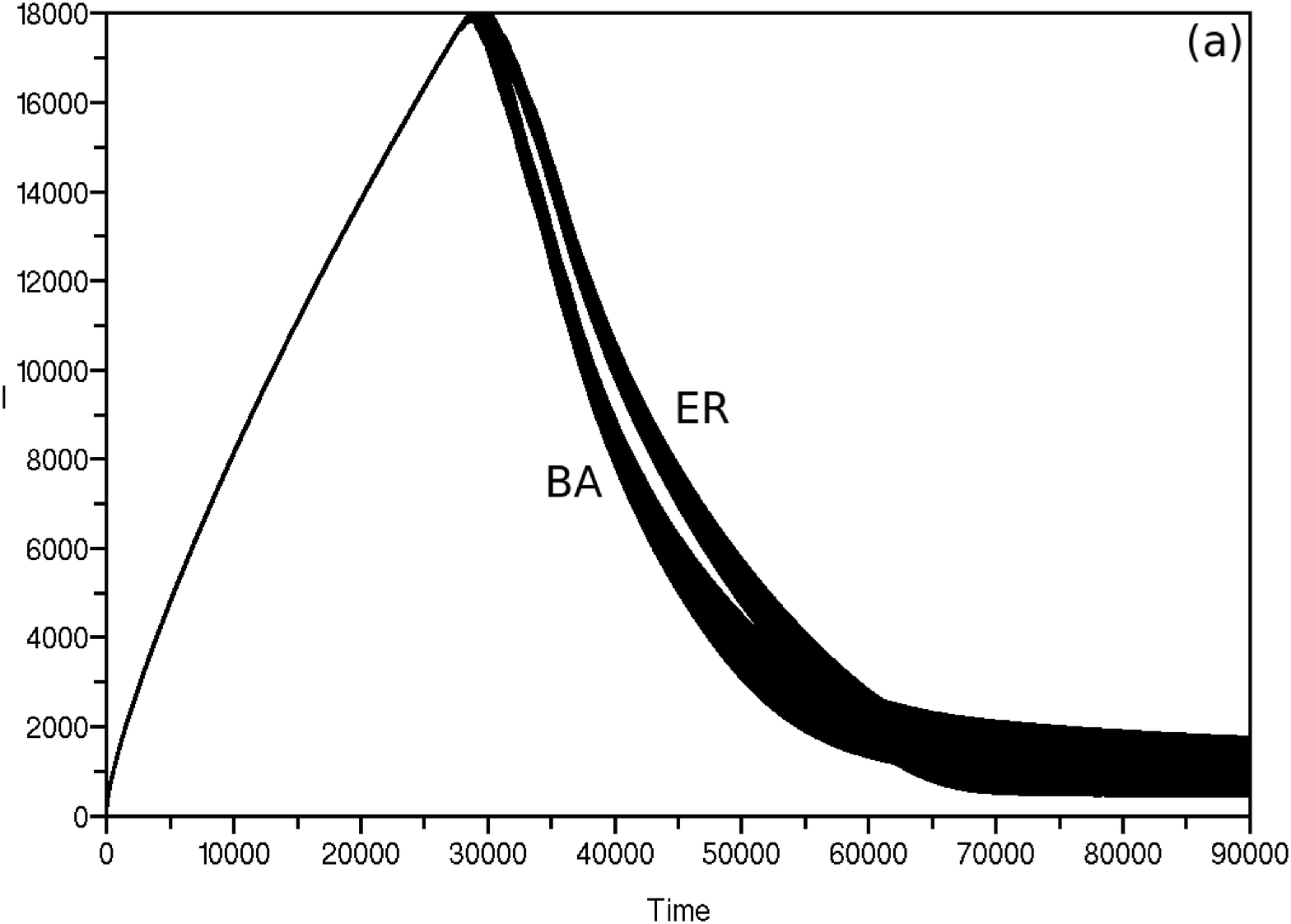}
  \includegraphics[scale=0.2]{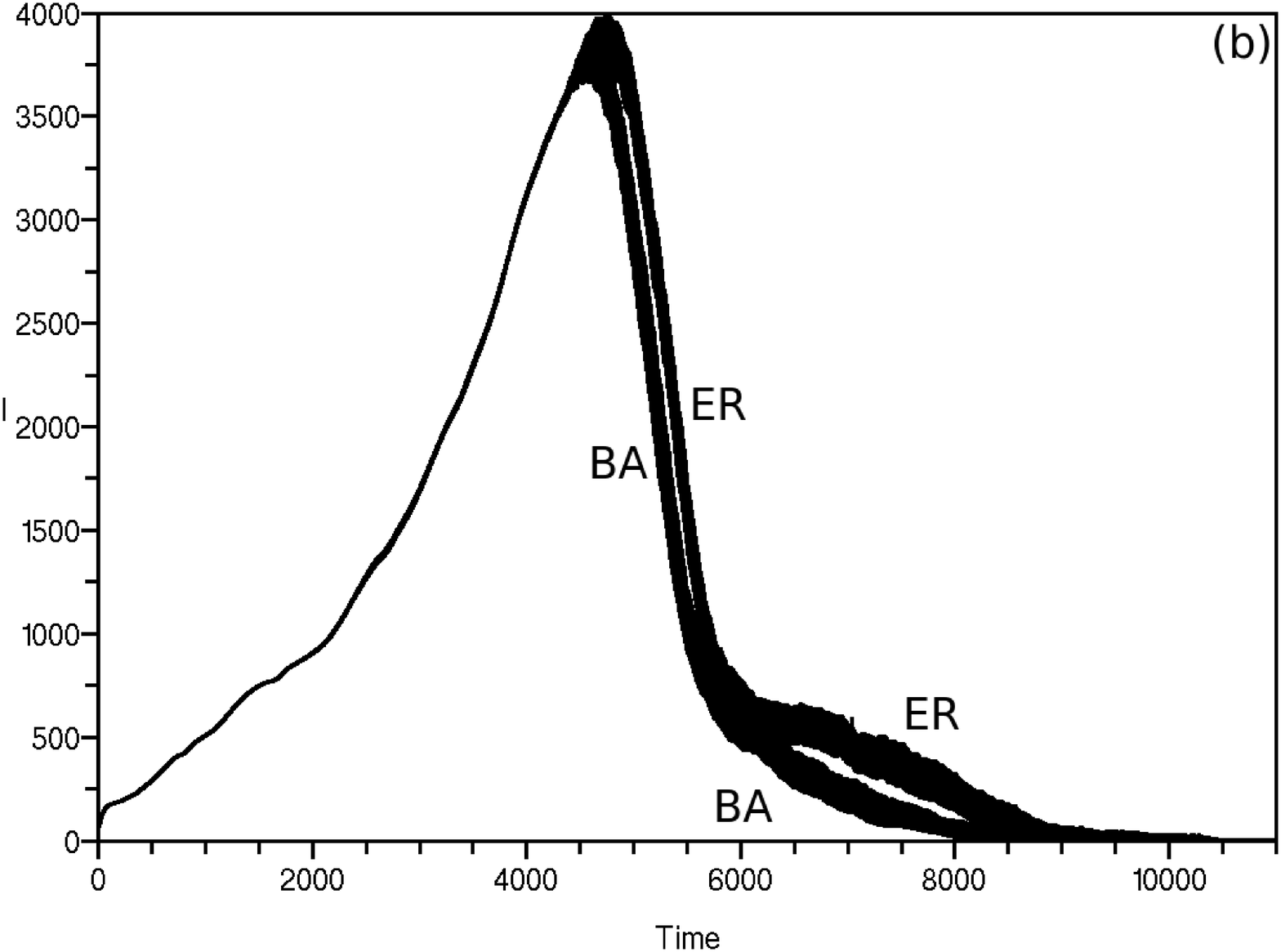}
  \caption{The amount of disease $I$ in the mesh ($y-axis$) along time
  ($x-axis$). (a) Fick diffusion model and (b) Gray-Scott model. \textit{Two
  sources}, $N=500$ and $\langle k\rangle \approx 6$ configuration. The
  standard deviation is shown at one fifth of its real value.}  \label{fig:09}
  \end{center}
\end{figure}

We also investigated the evolution of the disease when two sources
were established as initial conditions. Once again, a total of $100$
realizations was considered for each parameter configuration. The
resulting shape of the curves was similar to that observed for the
\textit{one-source} configuration. Naturally, the Fick diffusion with two
symmetrically displaced sources resulted in faster increase in the
total amount of disease, so that the threshold was quickly overcame
(about $20000$ time units before the \textit{one-source} case -
fig.~\ref{fig:07}-a, fig.~\ref{fig:08}-a and fig.~\ref{fig:09}-a).
The main strategy against the Fick diffusion is to enclose the
sources, which was obtained as soon as the chain effect was
triggered. Afterwards, the node only had to keep generating
antidote in order to completely eliminate the already spread
disease. A disease decrease rate similar to the \textit{one-source}
configuration was also expected. Once many node were still in their
original positions, there were many susceptible nodes to be shared
between the sources. Another interesting effect occurred when one of
the sources engaged all of the available node. The control of the
second source turned out to be indirect, \textit{i.e.} due to the
antidote generated by the node activated by the first source. The
latter effect slowed down the elimination of the disease and increased
its minimal level along the last steps, since the antidote could not
reach the source as effectively as could be achieved by node
displacement. This effect implied higher standard deviation of the
disease intensity $I$, specially at its minimal levels.

The configuration with higher number of nodes and connectivity (fig.~\ref{fig:08}-a and fig.~\ref{fig:09}-a) resulted on decrease of the efficiency in the BA network in the last stage of the defense dynamics. The uniform distribution of ER connections resulted on average in a higher efficiency in the enclosement of both sources. Over time, ER better managed the swapping of node between both sources. On the other hand, \textit{hubs} requests resulted in a higher concentration of node around one of the sources (\textit{e.g.}, source $1$). Consequently, the node were hardly activated due to the disease generated by the other source (\textit{e.g.}, source $2$). The latter effect diminished the elimination rate of the disease because one of the sources (\textit{e.g.}, source $2$) turned out to be indirectly controled, \textit{i.e.}, through the antidote generated only by the node which were activated by the first source (\textit{e.g.}, source $1$).

The initial effect of the two sources in the Gray-Scott model was the
creation of two large infected areas on both sides of the wall of
nodes. Each of them had approximately the same size as the area
generated by the \textit{one-source} configuration. The total amount of disease
before the first activation was nearly twice as much in the
\textit{two-sources} configuration than observed for the one-source
case. Consequently, when the spots and stripes reached the node, they
initially had a larger amount of disease to eliminate. The same three
stages were identified (fig.~\ref{fig:07}-b) as in the \textit{one-source}
case. However, the uniform distribution of neigbors in the ER network
favored a better distribution of the node among the many infected areas.  
This effect contributed to improve the defense ability of the
network and enhanced its efficiency. On the other hand, \textit{hubs}
made massive attacks against large infected areas. However, they
requested many node which were defending other areas. The
same effect contributed to the appearance of the second peak in
figure~\ref{fig:04}-b. The increase of network node
(fig.~\ref{fig:07}-b) resulted in better control of the disease
constrained by the ER network. In fact, the second peak (relapse) was
absent in this case. Finally, the increase in the connectivity of the
network, resulted in even faster elimination of the disease. In the
average, each request engaged more node, which contributed to the
steady reduction of the amount of disease.

\section{Conclusions}

Many natural phenomena involve interactions between two or more independent \textit{sub-systems} with specific properties (\textit{e.g.}, firemen combating forest fire, infection spreading into a healthy tissue while interacting with defensive cells, cleaners controlling oil spilling, pest control, \textit{etc}). The structure of each \textit{sub-system} can be modeled in terms of a network while the dynamics is represented by processes occuring in each network (\textit{e.g.}, the movement of agents or pattern formation). An interaction rule couple both \textit{sub-systems} in such a way that the evolution of one \textit{sub-system} depends on the other one and \textit{vice-versa}. Since the connections are responsible for the way the defensive agents communicate, they play a fundamental role in the behavior of such complex systems, \textit{i.e.}. they control the dynamical evolution of the agents (\textit{i.e}, node) which in turn, constrains the evolution of the dynamical pattern.  For example, the specific way in which groups of firemen are organized determines whether they will control or not the fire spreading. Similarly, the signal connectivity of anti-bodies (\textit{i.e.}, complex network) is crucial to efficiently activate them to stop an infection diffusion through a healthy tissue.

To investigate such phenomena, we proposed a dynamical hybrid system
composed of a regular and a complex network. The complex network
represented connected defensive agents (\textit{i.e.}, node)
self-organizing to eliminate patterns evolving in the regular
network which in turn, represented the unwanted process. According
to the local pattern intensity, the node were activated to liberate an
opposite diffusion aiming to eliminate the pattern. Two pattern
growth models were considered: Fick diffusion and Gray-Scott
reaction-diffusion. The defensive agents were connected following
Erd\"os-R\'enyi and Barab\'asi-Albert models. Two types of initial
conditions were investigated: \textit{one-source} and \textit{two-sources}. The role
of the network structure was investigated by using three network
configurations: (i) $N=300$ and $\langle k\rangle \approx 4$ (ii)
$N=500$ and $\langle k\rangle \approx 4$ (iii) $N=500$ and $\langle
k\rangle \approx 6$.

The main results included the better performance obtained by the BA
comparatively to the ER network to any chosen configuration. The
\textit{hub}-based characteristic of the BA network provided massive
attacks against the disease. Heavy defense was crucial in the
beginning in order to fast accelerate the ratio of decrease of the amount of disease in the regular
network. These massive attacks avoided much leakage and emergence of
isolated patterns which were present at higher rates in the ER case.
Isolated patterns were responsible for the relapse of the disease. The
increase in the number of network nodes and in their connectivity
contributed significantly to faster eliminate the disease. These results have shown the importance of \textit{hubs} in defensive networks. \textit{Hubs} contribute to diminish the \textit{average path length} in the network. Consequently, on average the hierarchical level with maximum number of nodes can be reached earlier in the BA than in the ER network. As a result, a more effective defense can be evaluated when the disease is concentrated in a large area. On the other side, despite of the better efficiency of the BA network, the uniform distribution of nodes in the ER network contributed to efficient defense strategies when many isolated patterns emerged on different places in the regular network.

Future developments include: (i) investigation of optimal network structure 
to efficiently eliminate the pattern, (ii) analysis of how the system 
properties scale with its size, (iii) study of the pattern evolution under network perturbations (\textit{e.g.}, node attack or edge rewiring), and (iv) improvement of the model by inclusion of other communication protocols taking 
place in the defensive network, such as broadcasting.

\begin{acknowledgments}

LECR is grateful to CNPq for financial support. LFC is grateful to
CNPq (308231/03-1) and FAPESP (05/00587-5) for financial support.

\end{acknowledgments}

\bibliography{pre_grayscott}

\end{document}